\documentclass[a4paper,11pt]{article}
\pdfoutput=1 

\usepackage{jinstpub} 

\usepackage{lineno}
\title{\boldmath Measurements of an AC-LGAD strip sensor with a 120 GeV proton beam}

\usepackage{siunitx} 
\DeclareSIUnit\sq{\ensuremath{\Box}}                           

\usepackage[rawfloats=true]{floatrow}

\usepackage{subcaption}
\makeatletter
\renewcommand\p@subfigure{\thefigure.}                        
\makeatother

\usepackage{xcolor}

\author[1]{Artur Apresyan,}
\author[2]{Wei Chen,}
\author[2]{Gabriele D'Amen,}
\author[1]{Karri F. Di Petrillo,}
\author[2]{Gabriele Giacomini,}
\author[1]{Ryan E. Heller,}
\author[3]{Hakseong Lee,}
\author[3]{Chang-Seong Moon,}
\author[2]{Alessandro Tricoli}

\affiliation[1]{Fermi National Accelerator Laboratory, \\PO Box 500, Batavia IL 60510-5011, USA}
\affiliation[2]{Brookhaven National Laboratory,\\Upton, 11973, NY, USA}
\affiliation[3]{Kyungpook National University,\\Daegu, South Korea}

\abstract{
The development of detectors that provide high resolution in four dimensions has attracted wide-spread interest in the scientific community for several applications in high-energy physics, nuclear physics, medical imaging, mass spectroscopy as well as quantum information.  
LGAD silicon sensors with finely segmented AC-coupled electrodes can provide precise spatial and timing measurements of incident minimum ionizing particles. Such AC-coupled LGADs, also known as AC-LGADs, are therefore considered as candidates for future detectors to provide 4-dimensional measurements in a single sensing device with 100\% fill factor. This article presents the first characterization of an AC-LGAD sensor with a proton beam of \SI{120}{\GeV} momentum at Fermilab. The sensor consists of strips with \SI{80}{\micro \m} width, fabricated at Brookhaven National Laboratory. The signal properties, efficiency, spatial, and time resolution are presented. The experimental results show that the time resolution of such an AC-LGAD is compatible to standard LGADs with similar gain, and that AC-LGADs can be segmented with fine pitches as standard strip or pixel detectors.
}

\keywords{Solid state detectors; Timing detectors; Particle tracking detectors (Solid-state detectors); Si microstrip and pad detectors.}

\begin{document}
\maketitle
\flushbottom


\section{Introduction}

The development of high spatial resolution pixel or strip detectors with high per-pixel or per-strip time resolution has been one of the major technological drivers in collider physics in recent years. This need results from challenges posed by future experiments at particle colliders, including high interaction rates, high particle production densities, large backgrounds from multiple interactions per bunch crossing (pileup), fast readout speed, and significant particle momentum smearing due to beam dispersion.

Current particle trackers in collider experiments are based on silicon technology with a spatial resolution of few tens of microns, while novel silicon technologies have recently allowed timing resolution of few tens of picoseconds, for instance with Low Gain Avalanche Detectors (LGADs)~\cite{GIACOMINI201952}~\cite{micronlgad}. The development of LGAD technology was prompted by the need to time resolve the tremendous number of particle tracks emerging from  the interaction regions in high energy physics experiments, and to improve the reconstruction accuracy of the primary particle interaction in high pileup conditions~\cite{hartmuth}. The ATLAS and CMS experiments~\cite{CMS:2667167,Collaboration:2623663} at the High Luminosity LHC (HL-LHC)~\cite{CERN-ACC-2015-0140,CERN-ATS-2012-236}, which is expected to begin in 2026, have developed fast-timing detectors based on LGAD sensors.

The LGAD is based on a simple $p$--$n$ diode concept, where the diode is fabricated on a thin high-resistivity $p$--type silicon substrate. A highly-doped $p^{+}$--layer is placed between a highly-doped $n^{++}$ implant and the $p$--type bulk. This $p^{+}$--layer is also known as the "gain" layer. The application of a reverse bias voltage creates an intense electric field in this superficial region of the sensor, able to start an avalanche multiplication for the electrons. The gain is limited to a factor of typically 10-100, such that the noise is kept low, in comparison to the case of avalanche photodiodes. The drift of the multiplied carriers through the thin substrate generates a fast signal with a time resolution of few tens of picoseconds.

The LGAD sensors developed for the ATLAS and the CMS timing-detectors have relatively large pads of about $1.3 \times 1.3~\rm{mm}^2$ size and a substrate thickness of \SI{50}{\micro \m}. The pad dimensions in these detectors are designed to be far larger than the substrate thickness in order to achieve a more uniform electric field. With larger pad sizes, for most positions on the sensor, the electric field can be approximated as a parallel plate capacitor, and edge-effects can be neglected. These large pads therefore allow a uniform multiplication to occur over the entire surface of the sensor. Recent research has focused on how to segment LGAD sensors~\cite{RSD_NIM} while maintaining the fine LGAD time resolution, e.g. with pixels or strips that have pitches in the tens of microns in order to achieve fine spatial resolution without the limitations of reduced fill-factor. Several designs have been proposed, for example silicon microstrip sensors that implement the gain layer under the strip~\cite{WADA2019380} and double-sided sensors that feature a large uniform gain layer on the opposite side of the patterned electrodes~\cite{DALLABETTA2015154, PELLEGRINI201624}. Both types of designs present challenges, more specifically the sensors in the first category have multiplication only in the active area close to the center of the strip, while those in the second category have compromised timing properties because of the thicker substrate, i.e. \SI{200}-\SI{300}{\micro \m}. Furthermore, one of the key features in LGAD sensors is the presence of Junction Termination Edges (JTE's) at the cell border, which are required to avoid premature breakdowns at the cell edges. The presence of JTE's and the gap itself between LGAD cells lead to an inherent limitation that 100\% fill factor cannot be achieved.  The new technology of AC-coupled LGADs (AC-LGADs~\cite{RSD_NIM}) has been demonstrated~\cite{ACLGADprocess,8846722} as a good candidate for a 4-dimensional (4-D) silicon detector to provide time resolution in the few tens of picoseconds and segmentation of few tens of microns with a fill factor of 100\%. Such a 4-D detector could find important applications in various fields in addition to high-energy physics.

This article reports the results of the first test beam of an AC-LGAD, using a strip sensor fabricated  at Brookhaven National Laboratory (BNL) and the \SI{120}{\GeV} proton beam at the Fermilab Test Beam Facility (FTBF). Section~\ref{sec:sensor} presents a description of the AC-LGAD sensor. Simulations of AC-LGAD signals are presented in Section~\ref{sec:simulation}, and the experimental setup at the FTBF is described in Section~\ref{sec:setup}. Experimental results are presented in Section~\ref{sec:results}. These results include measurements of sensor signal properties, including the induced signal on adjacent strips, measurements of the device efficiency, and measurements of spatial and time resolution. Conclusions and outlook are presented in Section~\ref{sec:conclusions}.

\section{The AC-LGAD sensor}\label{sec:sensor}

\begin{figure}[h!]
\centering 
\includegraphics[width=.65\textwidth]{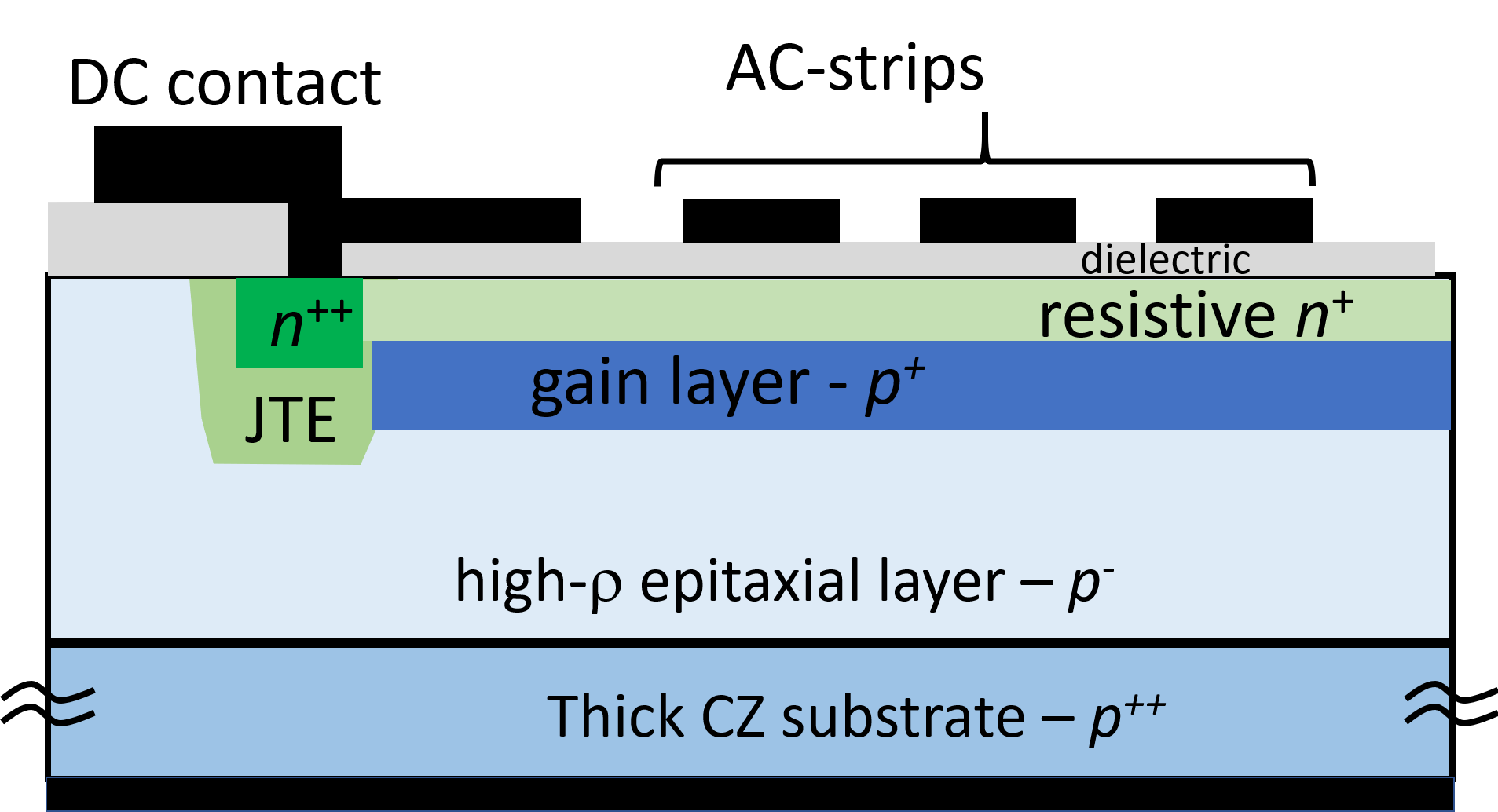}
\caption{\label{fig:sketch}  Cross section of a segmented AC-LGAD.  For simplicity, only three AC electrodes are shown, and the figure is not to scale.}
\end{figure}

The AC-LGAD sensor studied in this article was fabricated at BNL, using a class-100 silicon processing facility, following the process outlined in Ref.~\cite{ACLGADprocess}. Figure~\ref{fig:sketch} shows the cross-section of a segmented AC-LGAD sensor. 

The substrate is a \SI{50}{\micro \m} thick, $p^-$ -- type epitaxial layer. 
Unlike a standard DC-coupled LGAD, the $n^+$ layer was implanted with a very low dose of phosphorus, in order to increase the layer's resistance~\cite{GIACOMINI201952}. Just below the $n^+$ layer, the gain layer was implanted with boron, with a dose chosen to result in a breakdown voltage of about \SI{-200}{\V}, based on simulations. 

From experimental measurements at BNL the depletion voltage of the wafer, from which the sensor was cut, was measured to be approximately \SI{-150}{\V}, while the breakdown voltage was measured to be approximately \SI{-220}{\V}. Operating  voltages for this wafer are thus chosen in the range between the depletion voltage and breakdown voltage, i.e. between \SI{-150}{\V} and \SI{-210}{\V}. The depletion voltage for the LGAD is typically higher than the voltage needed to deplete a simple junction fabricated onto the same substrate. The application of additional voltage is required to fully deplete the gain layer. 

At the edge of the active area, defined horizontally by the $n^+$ and the gain layer sheets, a high-dose $n^{++}$ -- type implant was inserted to make a DC contact with the external world, and to provide the ground to the structure. This implant is inserted within a JTE. The JTE consists of a deep and low-dose phosphorus implant that terminates the structure, and prevents the formation of high electric fields at the very edge of the active area. 

Above the active area, a thin dielectric layer of about \SI{100}{\nano \m} of silicon  nitride was deposited. On top of this layer, metal pads are placed to define the AC-coupled electrodes of the structure. These pads are connected to the read-out electronics.

Outside the implanted areas, there is the thick thermal oxide, that never underwent an etching step, covered by the thin layer of silicon nitride.

\begin{figure}[h!]
    \centering 
    \includegraphics[width=0.7\linewidth, height=5cm, keepaspectratio]{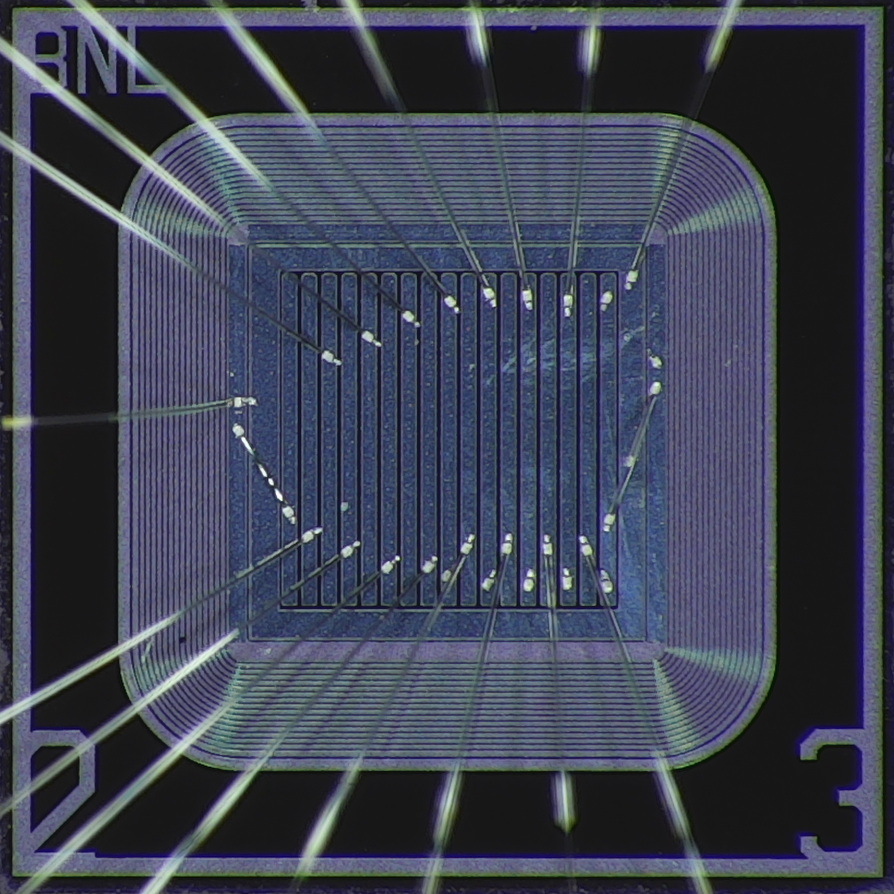}
    \caption{Photograph of the AC-LGAD sensor. The seventeen individual strips are referenced according to labels 0 to 16, from left to right. The innermost square ring adjacent to the strips is the DC contact. Strips 1 to 15, along with the DC pad, are wire-bonded to the 16-channel readout board. }
    \label{fig:AC-LGAD_BOARD}
\end{figure}

The sensor used in this study is shown in Fig.~\ref{fig:AC-LGAD_BOARD}, and it was cut out from a $4$-inch wafer, populated with several AC-LGAD structures that differed in the dimensions of their active areas and the patterning of their metal pads. The sensor active area is $2 \times 2~\rm{mm^2}$, and consists of an array of $17$ AC-coupled metal strips surrounded by a DC-connected pad. The strips are \SI{1.67}{\milli \m} long with pitch of \SI{100}{\micro \m}. 

\begin{figure}[h!]
\centering 
\includegraphics[width=0.6\linewidth,height=1.8in]{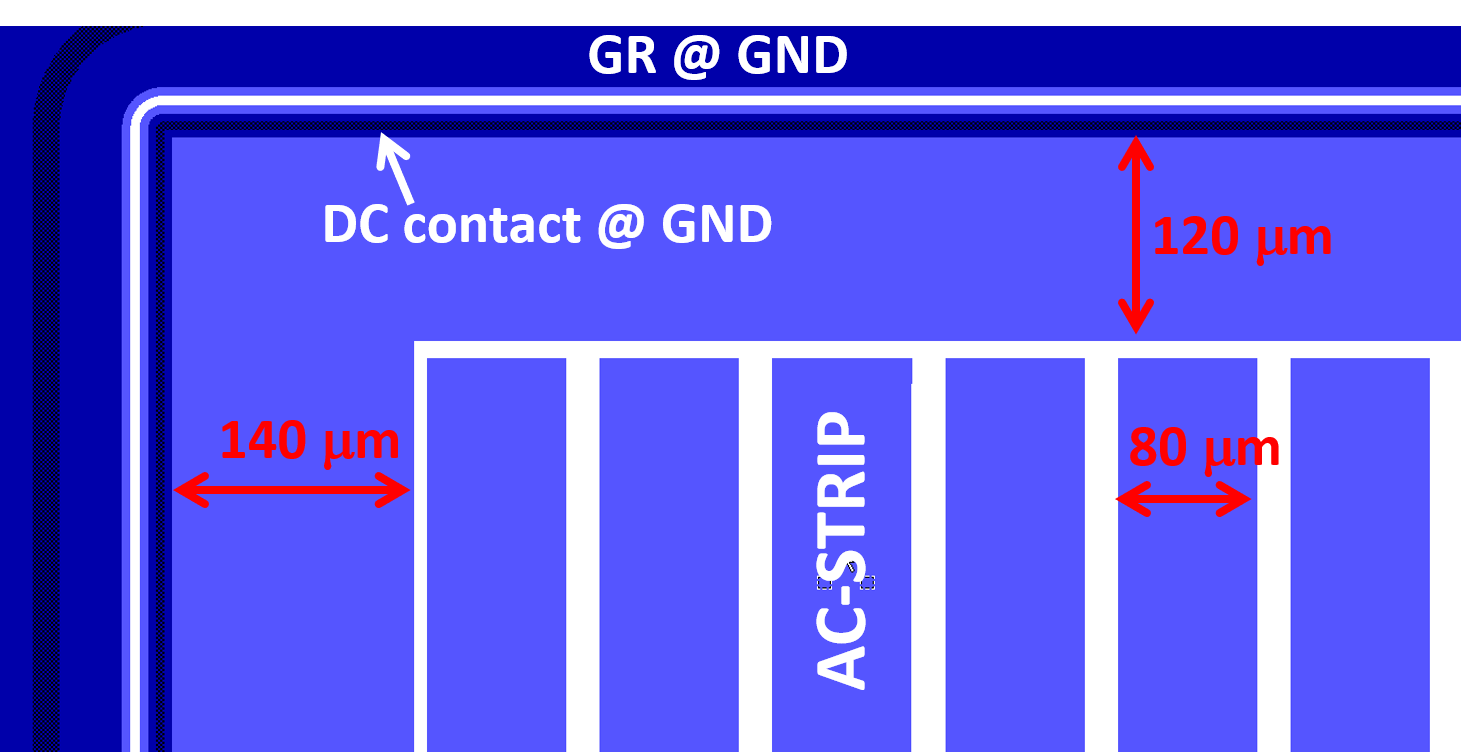}
\caption{ \label{fig:layout} The layout of a corner of the sensor is shown, with metallized regions shown in blue. The device strip width is \SI{80}{\micro\m}, the inter-pad gap is \SI{20}{\micro\m}, while the width of the DC-pad contacts varies between $120$ and \SI{140}{\micro \m}. A Guard Ring (GR), shown in dark blue, surrounds the DC-pad and both sets of contacts are referenced to ground (GND). 
}
\end{figure}

The layout of the sensor used in this study is shown in Figure~\ref{fig:layout}. Metallized regions are shown in blue. The width of each metallized electrode is \SI{80}{\micro \m}, resulting in an inter-strip gap of \SI{20}{\micro \m}. The DC-pad has a width of \SI{120}{\micro \m} on the upper and right edges of the sensor, and \SI{140}{\micro\m} on the lower and left sides. The DC contact between metal and silicon is made at the device's edges, and is embedded into the JTE. The gain layer and the $n$--resistive layer (i.e. the active area of the device) are located within the area delimited by the JTE. A Guard Ring (GR) surrounds the active area, and is grounded during operations.

\section{AC-LGAD simulations}\label{sec:simulation}

We simulated, by means of the TCAD numerical simulator SILVACO~\cite{silvaco}, an AC-LGAD structure with a geometry similar to that of the sensor under study. To limit the computational resources necessary for the simulation, we kept the number of the grid nodes low by simulating only a part of the active region of the device, starting from the edge at X~=~0 (where the DC contact is) up to a lateral  dimension of X = \SI{0.5}{\milli\m}
The strip pitch and inter-pad gap were the same as in the device under test, i.e. \SI{100}{\micro \m} and \SI{20}{\micro \m}, respectively. The doping profiles are qualitatively close to those of the sensor, and a gain on the order of $10$ is attained in the simulation.

\begin{figure}[h!]
\centering 
    \begin{subfigure}[h]{0.48\textwidth}
        \centering
        \includegraphics[width=\linewidth]{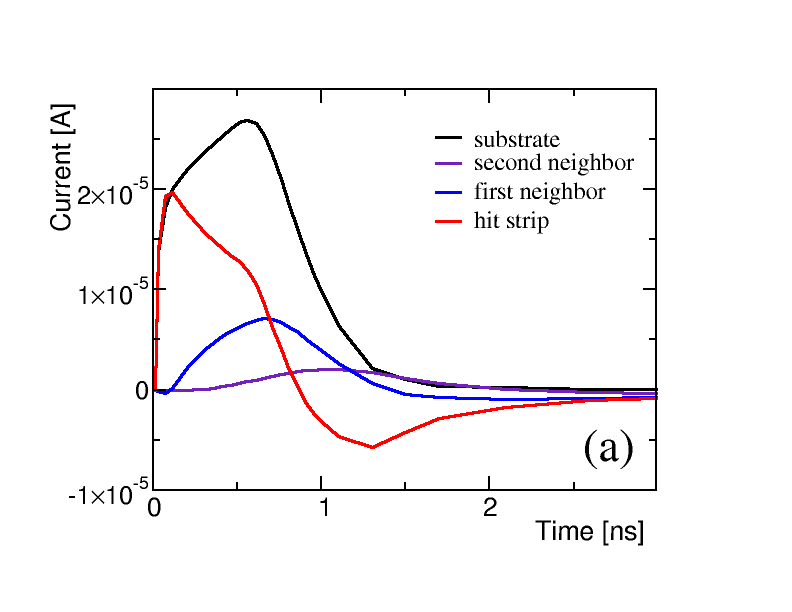}
        \caption{Signal simulation for AC-strips and substrate}\label{fig:simulation_strips}
    \end{subfigure}%
    ~ 
    \begin{subfigure}[h]{0.48\textwidth}
        \centering
        \includegraphics[width=\linewidth]{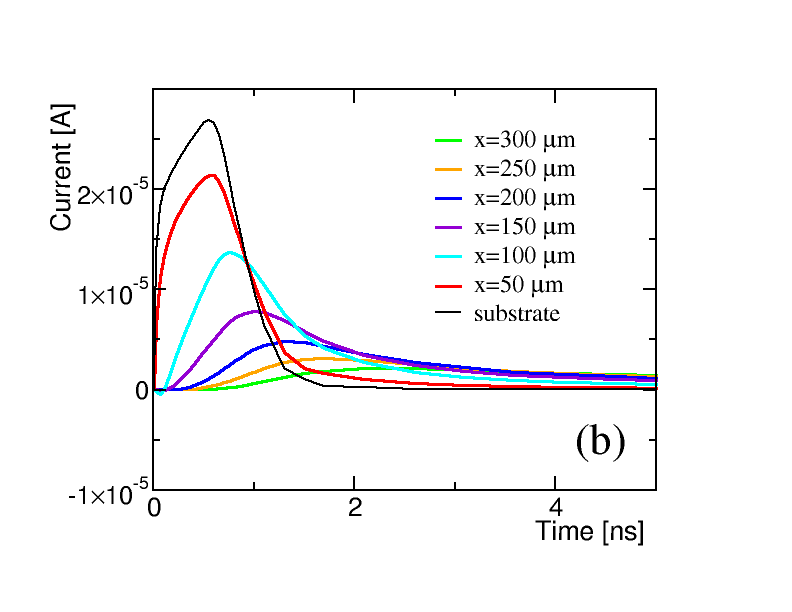}
        \caption{Signal simulation at the DC-pad contact }\label{fig:simulation_dc_pad}
    \end{subfigure}
    \caption{TCAD simulations of current pulses for an AC-LGAD sensor whose geometry matches the sensor under study: (a) bipolar current pulses on different strips generated by a MIP traversing the sensor at the middle of the "hit" strip; (b) unipolar current pulses from the DC-pad contact as a function of the incident particle's perpendicular distance, x, from the contact. 
    } 
    
\label{fig:simulation}
\end{figure}

Figure~\ref{fig:simulation_strips} shows the simulated current pulses at the metal electrodes, i.e. the AC-coupled strips, when a minimum ionizing particle (MIP) traverses the silicon vertically in the middle of a strip (labeled as "hit strip" in the figure). The simulated current pulses of the two first neighboring strips are also shown. As pulses from strips on opposite sides of the hit strip are very similar, only one of them is reported. The same applies for the second neighbors. 
The current pulse at the substrate is also shown. For comparison purposes, the pulses are shown with the same polarity.

By Kirchhoff's law, the sum of the currents  at all the electrodes (substrate, DC-pad contact and AC pads) is null. Since nearby electrodes experience a sharing of the total current, the current signal at any AC pad will have a smaller amplitude than the substrate signal. As a consequence, the signal collected by a single fine-pitched electrode will correspond to only a fraction of the full signal generated in the substrate.

Figure~\ref{fig:simulation_dc_pad}, shows current pulses at the DC-pad contact for different incident positions of a MIP. MIPs which traverse the active layer closer to the DC pad generate higher and shorter signals. Signals generated by MIPs which traverse the sensor further away from the DC pads are spread in time due to the high value of the RC time constant associated to the resistive $n$-layer and the AC pads. While the integral of these pulses is constant and equal to the generated charge multiplied by the gain, the amplitude of the smaller pulses may fall below the threshold of the read-out electronics.

\section{The experimental setup at the FNAL Test Beam Facility}\label{sec:setup}

Test beam measurements were performed at the FTBF~\cite{FTBF}, which provides a unique opportunity to characterize prototype detectors for collider experiments. A typical application is to place the device under test in the high energy beam, and measure its response to the beam particles passing through its active area. 

The FTBF provides a \SI{120}{\GeV} proton beam from the Fermilab Main Injector accelerator. The FTBF beam is resonantly extracted in a slow spill for each Main Injector cycle delivering a single \SI{4.2}{\s} long spill per minute, tuned to yield approximately 100,000 protons each spill. The primary beam of \SI{120}{\GeV} protons is bunched at \SI{53}{MHz}. The beam size is approximately $2$-\SI{3}{\milli \m} wide, though the intensity is peaked strongly in the center. All measurements presented in this paper were taken with such primary beam particles.

The AC-LGAD was wire-bonded to a $16$-channel readout board, designed at Fermilab. The $16$-channel board is designed to test sensors with sizes as large as $8.5\times8.5~\rm{mm}^2$ at voltages up to \SI{1}{\kilo \V}. Up to sixteen sensor outputs can be wire-bonded to the input pads of 2-stage amplifier chains based on the Mini-Circuits GALI-66+ integrated circuit. In this particular configuration amplifiers used a \SI{25}{\ohm} input impedance, an approximately \SI{5}{\kilo \ohm} total transimpedance, and a bandwidth of \SI{1}{\giga \hertz}.

The sensor and wire-bonding scheme for connection to the readout electronics is shown in Figure~\ref{fig:AC-LGAD_BOARD}. The two outermost metal strips were grounded by wire bond connection to the GR. The innermost fifteen strips were individually connected to independent inputs of the $16$-channel board. Strips are labeled $0$ to $16$. The DC pad at the border of the active area was connected to a readout channel too, in order to assess the gain of the sensor, as described in Section~\ref{DC_contact}.

A third stage of amplification was applied to signals from the AC strips, using the Mini-Circuits GALI-52+ evaluation board. The total transimpedance considering all three amplifier stages is approximately ~\SI{50}{\kilo \ohm}. The DC contact was read with only two stages of amplification, for a transimpedance of approximately ~\SI{5}{\kilo \ohm}.

The assembly of the AC-LGAD sensor and the $16$-channel read-out board was attached to an aluminum cooling block, and mounted on a remotely operated motorized stage inside an environmental chamber. A glycol-water solution was circulated through the cooling block, and kept the sensor at a constant temperature of $22 \pm 0.1$ $^{\circ}\rm{C}$. The relative humidity of the environmental chamber was kept to less than $10\%$. The sensor was biased to \SI{-210}{\V} for all results presented in this paper. The breakdown voltage of the sensor at this temperature is approximately \SI{-225}{\V}.

\begin{figure}[h!]
   \centering
    \begin{subfigure}[b]{0.9\linewidth}
      \center \includegraphics[width=\linewidth]{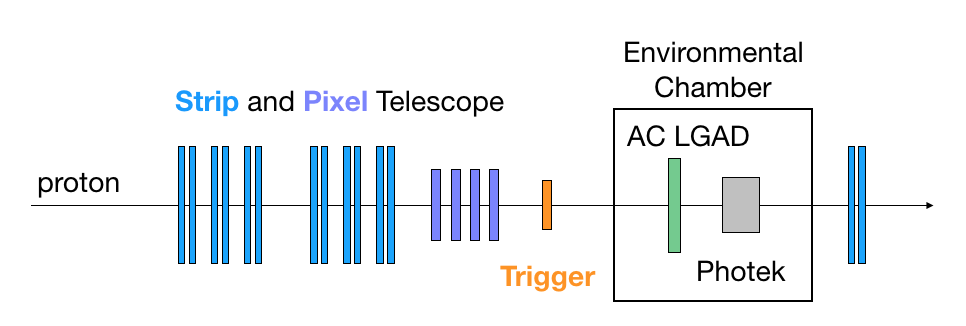}
      \caption{}
      \label{fig:FTBF_beam}
    \end{subfigure}
  \begin{subfigure}[b]{0.7\linewidth}
    \center \includegraphics[width=\linewidth]{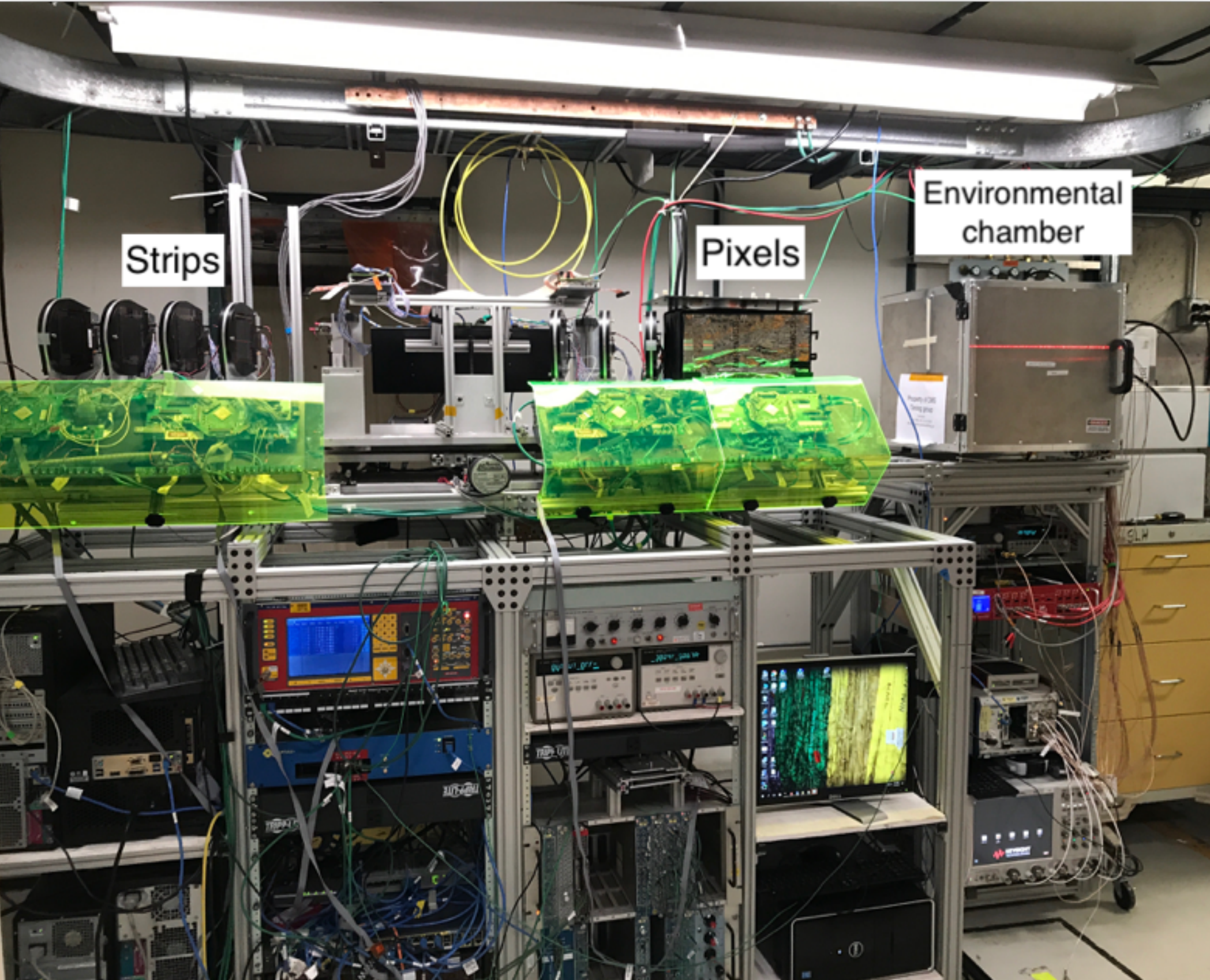}
    \caption{}
    \label{fig:FTBF}
  \end{subfigure}
\caption{(a) A schematic diagram of the test beam setup and FTBF telescope geometry. 
(b) A photograph of the experimental station and the telescope tracker at FTBF. In (b) the strip geometry is slightly modified from what is used in this result. }
\label{fig:TB_setup}
\end{figure}

The FTBF is equipped with a silicon tracking telescope to measure the position of each incident proton. The telescope consists of four pixel layers with cell size \SI[product-units = power]{100 x 150}{\micro \meter}, and fourteen strip modules with \SI{60}{\micro \meter} pitch, in alternating orientation along the x- and y- axes. The AC-LGAD was placed approximately \SI{2}{\meter} downstream from the center of the telescope. To detect and reject protons that scatter in any material along the beam line, two of the fourteen strip layers are located downstream of the environmental chamber. During this data-taking period, the spatial resolution of the telescope at the LGAD position was approximately \SI{50}{\micro \meter}, somewhat degraded due to the long extrapolation from the telescope and material along the beamline.

The telescope data acquisition hardware is based on the CAPTAN (Compact And Programmable daTa Acquisition Node) system developed at Fermilab. The CAPTAN is a flexible and versatile data acquisition system designed to meet the readout and control demands of a variety of pixel and strip detectors for high energy physics applications~\cite{4775101}. 

A Photek $240$~micro-channel plate (MCP-PMT) detector, placed inside the environmental chamber downstream from the AC-LGAD, was operated at $-3.7$~kV, and provided a precise reference timestamp. Its precision was previously measured to be smaller than \SI{10}{\pico \second}~\cite{RONZHIN2015288}. The resolution was confirmed to be better than \SI{10}{\pico \second} again in this experimental setup, by comparing timestamps from two Photek MCP-PMTs placed in the beamline at once.

The AC-LGAD and MCP-PMT waveforms were acquired using a Keysight MSOX92004A 4-channel oscilloscope, which provides digitized waveforms sampled at \SI{40}{GS/s}, and bandwidth limited to \SI{2}{\giga \hertz}. This oscilloscope's extremely deep memory is particularly well suited for the FTBF beam structure, allowing a burst of $50,000$ events to be acquired during each \SI{4.2}{\second} spill and written to disk during the longer inter-spill period.

The trigger signal to both the telescope and the oscilloscope originates in an independent scintillator coupled to a photomultiplier tube. The telescope and oscilloscope data are merged offline by matching trigger counters from each system. Events recorded by the two systems are kept synchronized by limiting the trigger rate to less than \SI{50}{\kilo \hertz}, limiting the recorded rate to ${\sim}50,000$ events per spill.  

A schematic of the experimental setup is shown in Figure~\ref{fig:FTBF_beam}, which presents the arrangement of AC-LGAD sensor with respect to the telescope tracker and triggers. Figure~\ref{fig:FTBF} shows a photograph of the setup.

\section{Experimental results}\label{sec:results}       

In this section, we present a number of studies performed on the AC-LGAD strip sensor in the FTBF facility. These studies include measurements of sensor signal properties, individual strip and total device signal collection efficiency, and characterization of the spatial and time resolution. A brief overview of the analysis strategy is presented below, and subsequent sections provide the details and results of each study. 

Events were required to have an incident proton with a position consistent with the sensor active area. This requirement was made by selecting events which have exactly one high-quality track in the FTBF telescope. Each track must have hits in at least sixteen, out of eighteen total, planes of the FTBF telescope. At least four of those sixteen hits must be in a pixel layer, and hits are additionally required in the two strip layers located downstream of the environmental chamber. Requirements were placed on the residual between the track and hit positions in the downstream layers in order to ensure that the track position was well measured, and that the incident proton did not scatter significantly inside the environmental chamber. Events were required to have a signal in the Photek MCP-PMT consistent with a MIP signal amplitude.

The coordinate system is defined such that the AC-LGAD strip length orientation is along the y-axis, and perpendicular to the x-axis. The orientation of the strips in the FTBF telescope's x-direction is reflected with respect to Figure~\ref{fig:AC-LGAD_BOARD}, while the y-orientation is unchanged.  

Since the oscilloscope has four channels, only data from the Photek and three of the sensor's channels could be collected simultaneously. To probe the properties of signals induced on neighboring strips over a range of distances, data were collected with varying combinations of adjacent and non-adjacent strips. In some of the following studies, data from more than three channels are stitched together using events from different readout configurations. In other studies, only events collected simultaneously from three adjacent strips are considered. 

An AC-LGAD strip is considered to have a hit in the following analysis if the signal's absolute amplitude is above \SI{110}{\milli \V}, in order to reject signals from noise. In order to study signal sharing between strips, we also define clusters of hits. Clusters are formed from adjacent strips which have a hit. As the DC-contact signals passed through only two rather than three stages of amplification, the DC hit threshold is correspondingly reduced to \SI{11}{\milli \V}.

Two methods are used to define the signal timestamp. The simplest method takes the time of the scope sample at which the signal reaches a maximum amplitude, or $t_{\rm{peak}}$. The second method, used for the time resolution measurement, performs a fit to the rising edge of the pulse to extract the time at which the pulse reaches $20\%$ of the maximum amplitude. This constant fraction timestamp is denoted as $t_{0}$. The reference time-stamp, $t_{\rm{ref}}$, is taken from the Photek signal, and defined with the same constant fraction method. 

\subsection{Strip signal properties}
\label{sec:signal_property}

\begin{figure}[h!]
\centering 
\includegraphics[width=.6\textwidth]{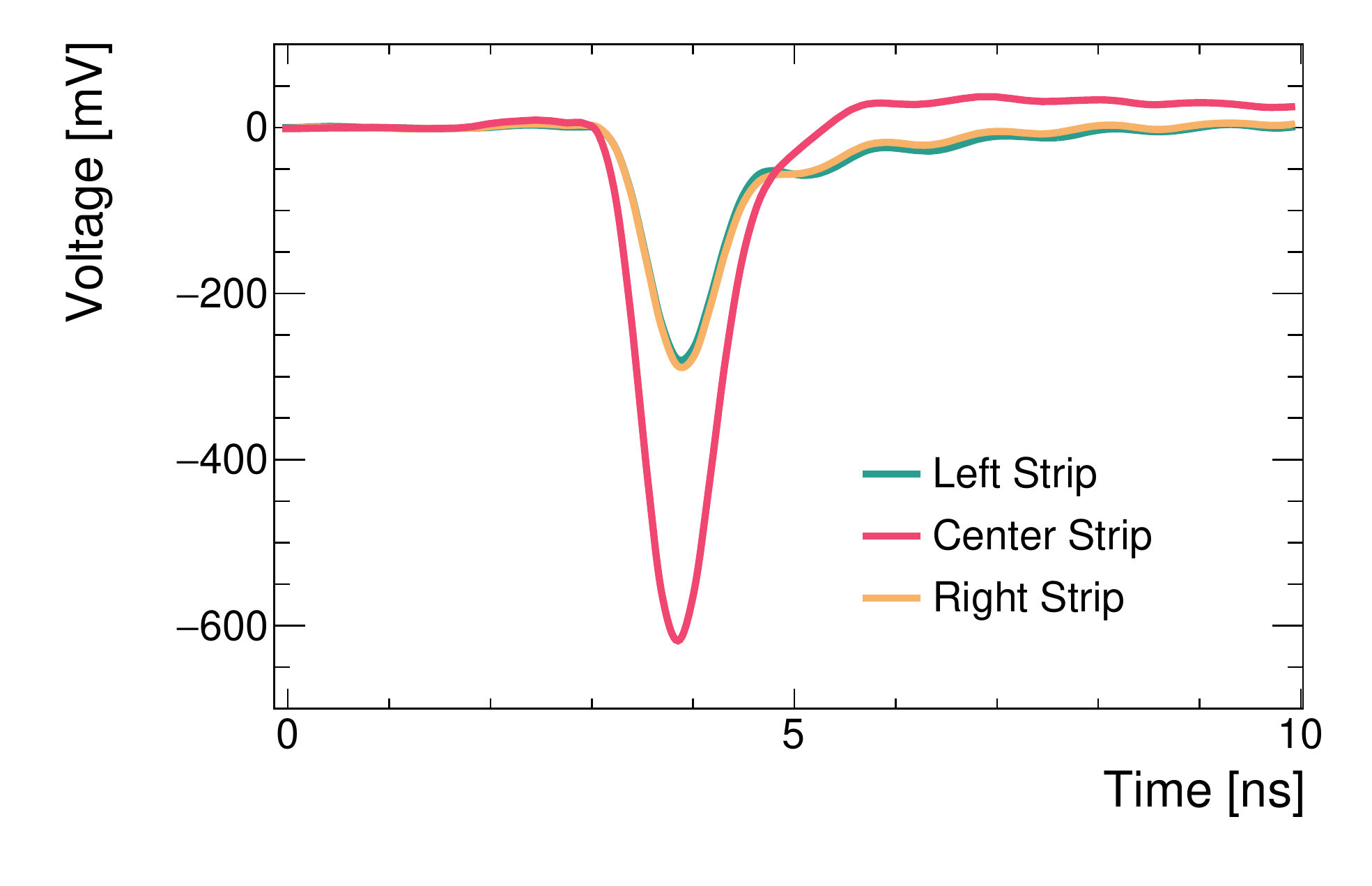}
\caption{Averaged waveforms for events in which three adjacent strips have signals with absolute amplitude values greater than \SI{110}{\milli \V}. }

\label{fig:signal_waveforms}
\end{figure}

Typical waveforms produced on AC strips by protons which traverse the sensor active area can be seen in Figure~\ref{fig:signal_waveforms}. Averaged waveforms are shown for events with a three-hit cluster, i.e. from clusters with signals in coincidence from at least three strips. Events are additionally required to have a proton with position $20.3 < x < 20.8$~mm and $22.8 < y < 24.2$~mm. For visualization purposes, the time of the signal is shown with respect to the reference timestamp. 
The center strip has an initial negative pulse with a duration of ${\sim}$\SI{1}{\nano\s} FWHM, followed by an overshoot.
The adjacent strips have lower amplitude pulses, with longer tails, qualitatively similar to the results of the simulation described in Section~\ref{sec:simulation} for current cross-talk between adjacent strips.

\begin{figure}[h!]
\centering 
    \begin{subfigure}[t]{0.4\textwidth}
        \centering
        \includegraphics[height=2.5in]{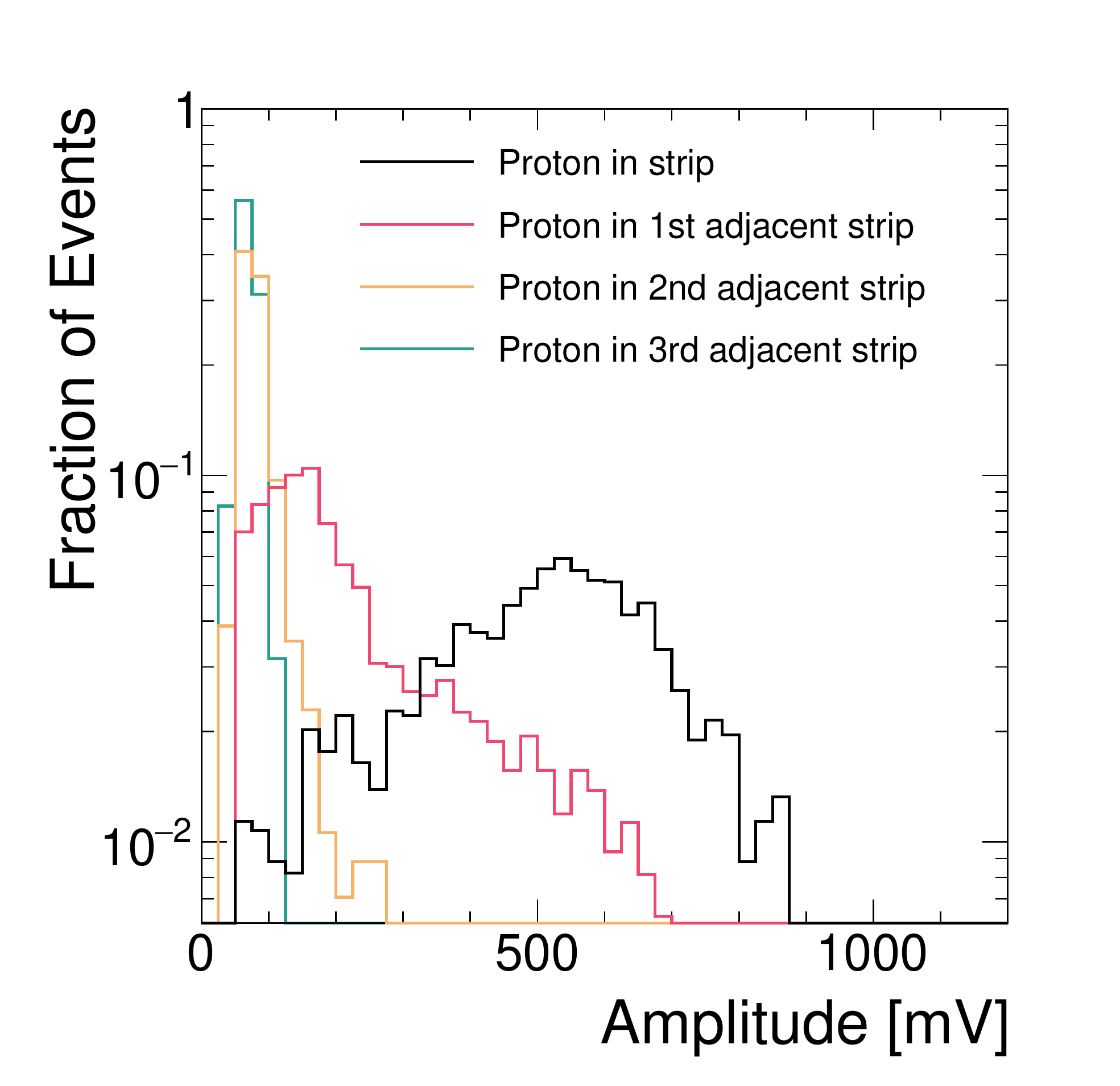}
        \caption{Strip amplitude distributions}\label{fig:amplitude_different_strips}
    \end{subfigure}%
    ~ 
    \begin{subfigure}[t]{0.55\textwidth}
        \centering
        \includegraphics[height=2.5in]{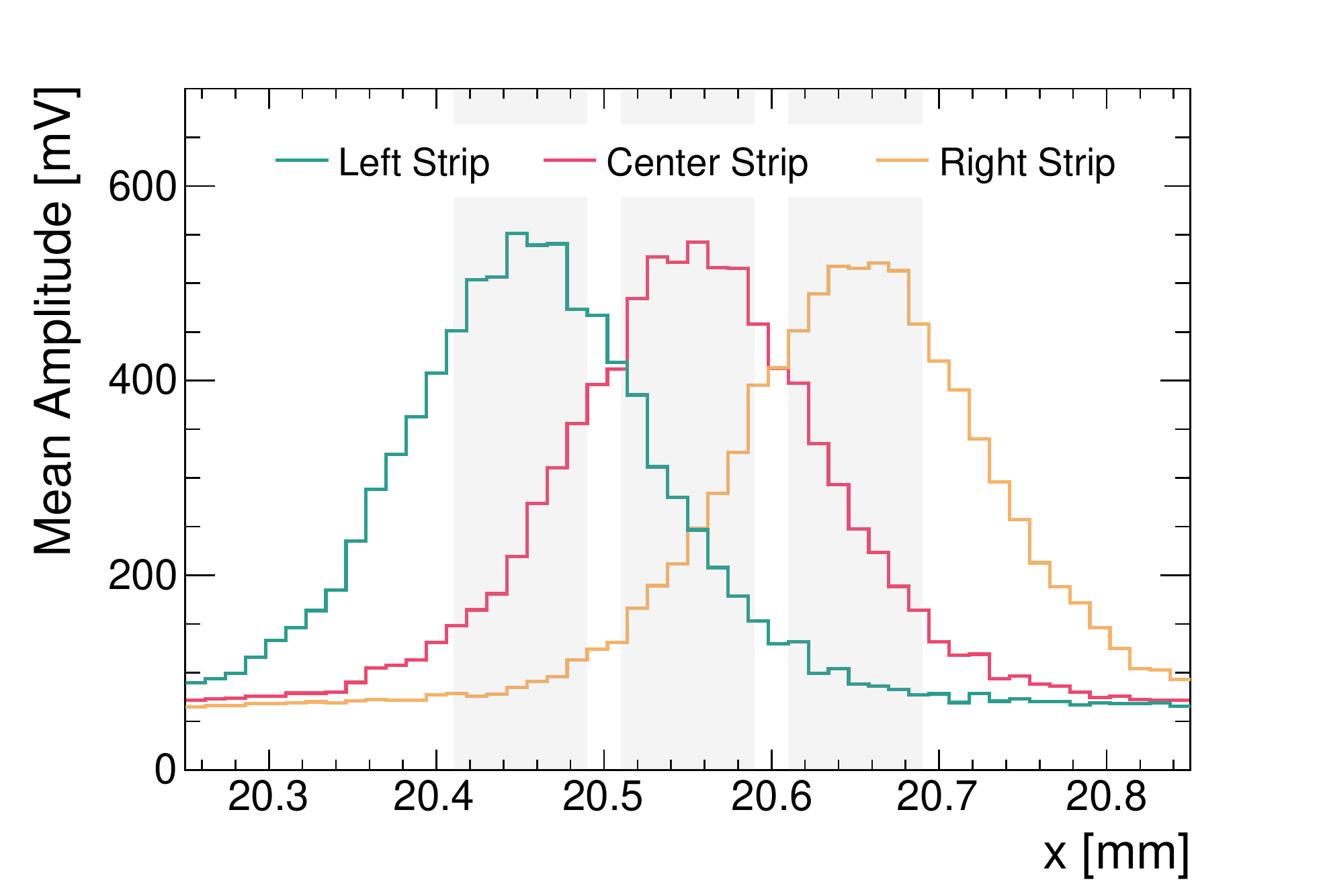}
        \caption{Mean strip amplitude versus proton position}\label{fig:amplitude_v_x}
    \end{subfigure}
    \caption{(a) Distributions of strip signal amplitudes in adjacent strips. The black histogram shows the amplitude of signals induced from protons which have positions consistent with the hit strip. The red, orange, and blue distributions show the amplitudes of signals induced on adjacent strips, progressively farther away from the hit strip and named as $1^{\rm st}$, $2^{\rm nd}$, and $3^{\rm rd}$ adjacent strip, respectively. Distributions are normalized in terms of fractions of events for comparison. (b) The mean amplitude of strip signals as a function of the incident proton track position in x. Shaded gray areas indicate the x-position of metallized strip regions.}

\label{fig:amplitude_v_position}
\end{figure}

The signal amplitude decreases with the strip's distance to the incident proton position, as shown in Fig.~\ref{fig:amplitude_v_position}. Figure~\ref{fig:amplitude_different_strips} shows the distributions of signal amplitude from a single strip, for protons incident to increasingly distant strips along the x direction. 
Figure~\ref{fig:amplitude_v_x} shows the mean amplitude of three channels, as a function of the incident proton position in the x-direction. Both figures demonstrate that the strip closest to the incident particle has the largest amplitude signal. Strips which are farther than the first adjacent strip to the primary strip are unlikely to produce a signal above threshold. The RMS noise is ${\sim}20$~mV, and for strips with the highest amplitude signal, the signal to noise ratio is $\rm{S} / \rm{N}{\sim}27$. The mean signal rise-time, or the time for the signal to rise from $10\%$ to $90\%$ of the maximum signal amplitude, is \SI{0.46}{\nano \s}.

Because data from only three channels could be recorded simultaneously, studies of amplitude distributions in Figures~\ref{fig:amplitude_different_strips} and~\ref{fig:amplitude_v_x} are used to infer typical cluster size. The majority of incident protons produce a cluster with three adjacent strips with signals above threshold. A small fraction of clusters is expected to produce a fourth or a fifth strip above threshold, and the signal amplitudes in these additional strips are small. We also infer that less than $1\%$ of clusters will have only one strip channel above threshold and roughly $20$-$30\%$ of clusters are estimated to contain two hits above threshold.

\begin{figure}[h!]
\centering 
    \begin{subfigure}[t]{0.48\textwidth}
    \centering 
    \includegraphics[height=2.4in]{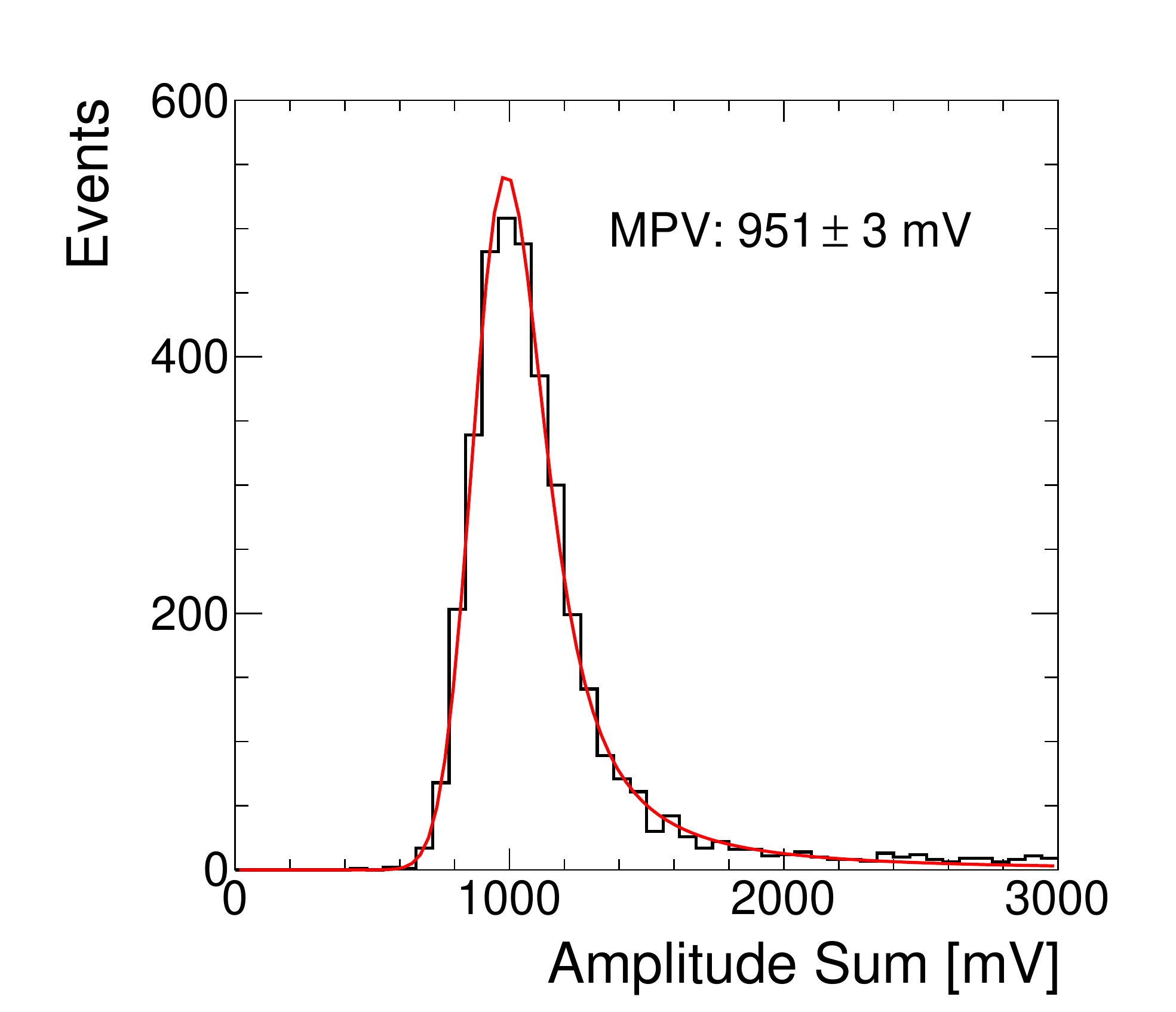}
    \caption{Sum of strip amplitudes}
    \label{fig:sum_amplitude}
    \end{subfigure}%
    ~ 
    \begin{subfigure}[t]{0.52\textwidth}
    \centering 
    \includegraphics[height=2.4in]{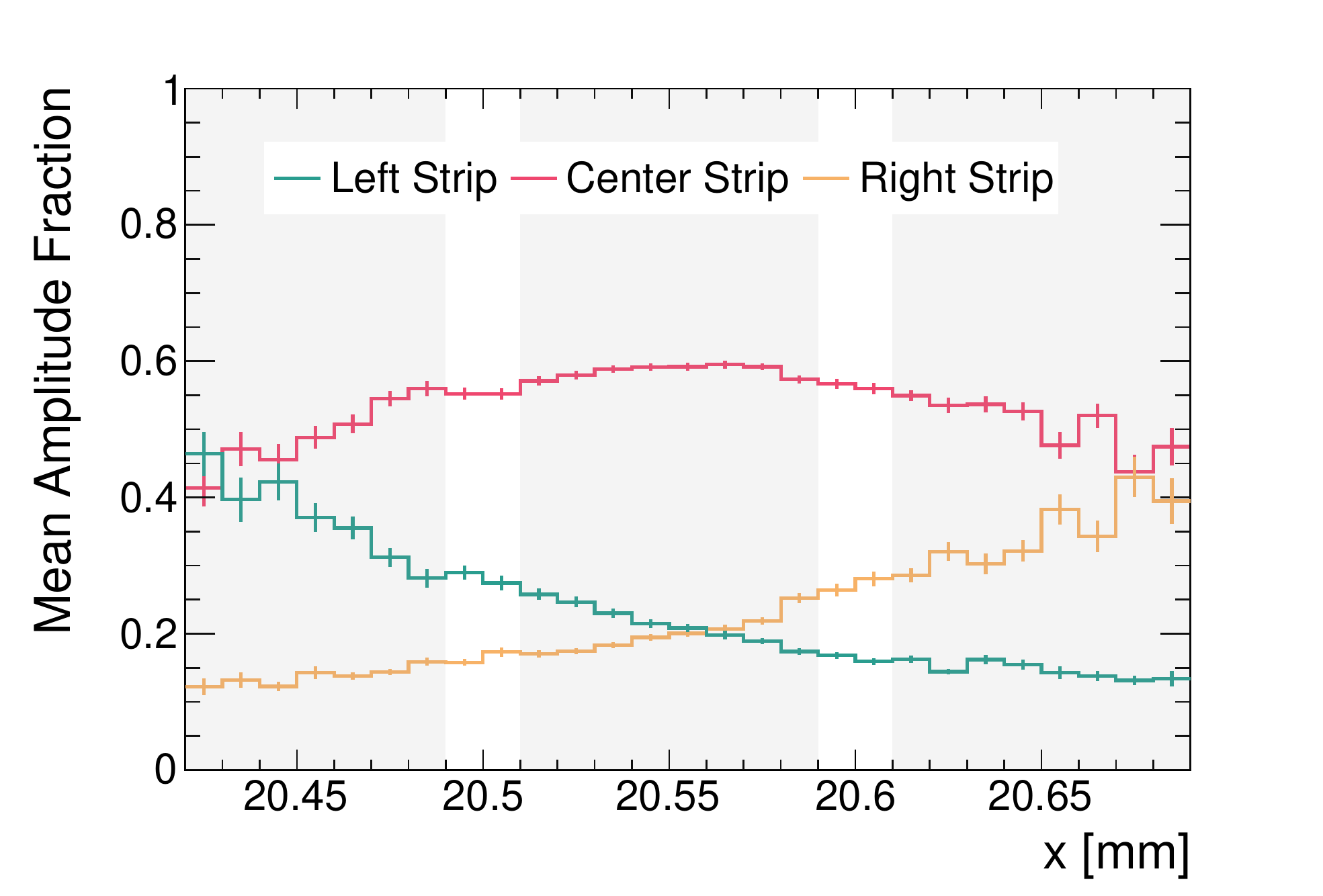}
    \caption{Relative strip amplitudes versus proton x}
    \label{fig:charge_fraction_v_x}
    \end{subfigure}
\caption{(a) Distribution of the sum of strip amplitudes for three-strip clusters. Data are shown in black, while a fit to data using a Landau function convoluted with a Gaussian function is shown in red. The most probable value (MPV) for the amplitude sum is shown on the figure, along with the statistical uncertainty from the fit. (b) The mean amplitude fraction for each strip in events with three-strip clusters. Data are shown for each channel as a function of the incident proton track x position. Shaded gray areas indicate the x-position of metallized strip regions. 
}
\label{fig:cluster_properties}
\end{figure}

The distribution of the sum of the signal amplitudes for clusters with three hits is shown in Figure~\ref{fig:sum_amplitude} together with a fit to data using a Landau function convoluted with a Gaussian function. The charge generated by a MIP is described by a Landau function, while the Gaussian accounts for the statistical variation of the gain and for the electronic noise. Since the cluster lateral dimension is mostly contained within three strips, it is expected that the sum of amplitudes of three strips in a cluster follows a Landau distribution, corresponding to the current signal generated in the substrate, as presented in Sec.~\ref{sec:simulation}.

The amplitude fractions of each strip within a cluster can be used to infer incident particle position and improve the spatial resolution in the measurement of the cluster centroid. The amplitude fraction is defined as the amplitude of a strip's signal, divided by the sum of hit amplitudes for strips which form a cluster. Figure~\ref{fig:charge_fraction_v_x} shows mean amplitude fractions for different strips in three-hit clusters as a function of the incident proton position. The majority (${\sim} 70\%$) of events included in Figure~\ref{fig:charge_fraction_v_x} have a proton incident on the center strip. Events with protons incident on the left or right strip of the cluster may also have produced signals in adjacent strips beyond those three shown in the figure, which are not read out in this configuration. The charge fractions in adjacent channels can be exploited to set constraints on the incident particle position.

\subsection{DC pad characterization}
\label{DC_contact}

Signals from the DC pad are characterized for events when the DC contact is directly struck by a proton. For these events, the RMS noise is ${\sim} 2$~mV, and the signal to noise ratio is $\mathrm{S}/\mathrm{N}{\sim}23$. The mean signal rise-time, defined as the time for the signalt o rise from $10\%$ to $90\%$ of the maximum signal amplitude, is \SI{0.46}{\nano \s}.   

\begin{figure}[h!]
\centering 
    \begin{subfigure}[t]{0.5\textwidth}
        \centering
        \includegraphics[height=2.5in]{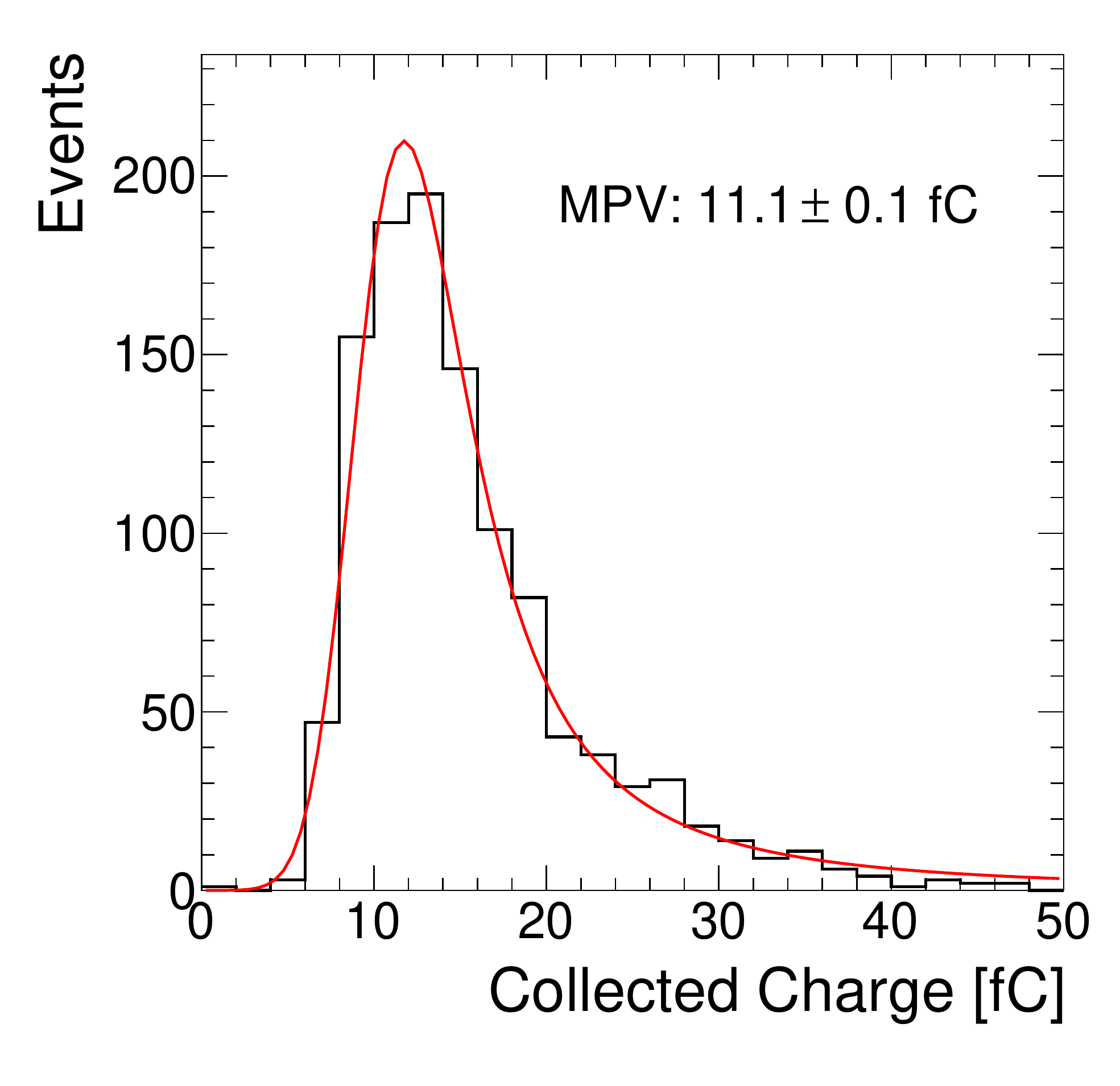}
        \caption{Charge collected by the DC pad}\label{fig:dc_landau}
    \end{subfigure}%
    ~ 
    \begin{subfigure}[t]{0.5\textwidth}
        \centering
        \includegraphics[height=2.6in]{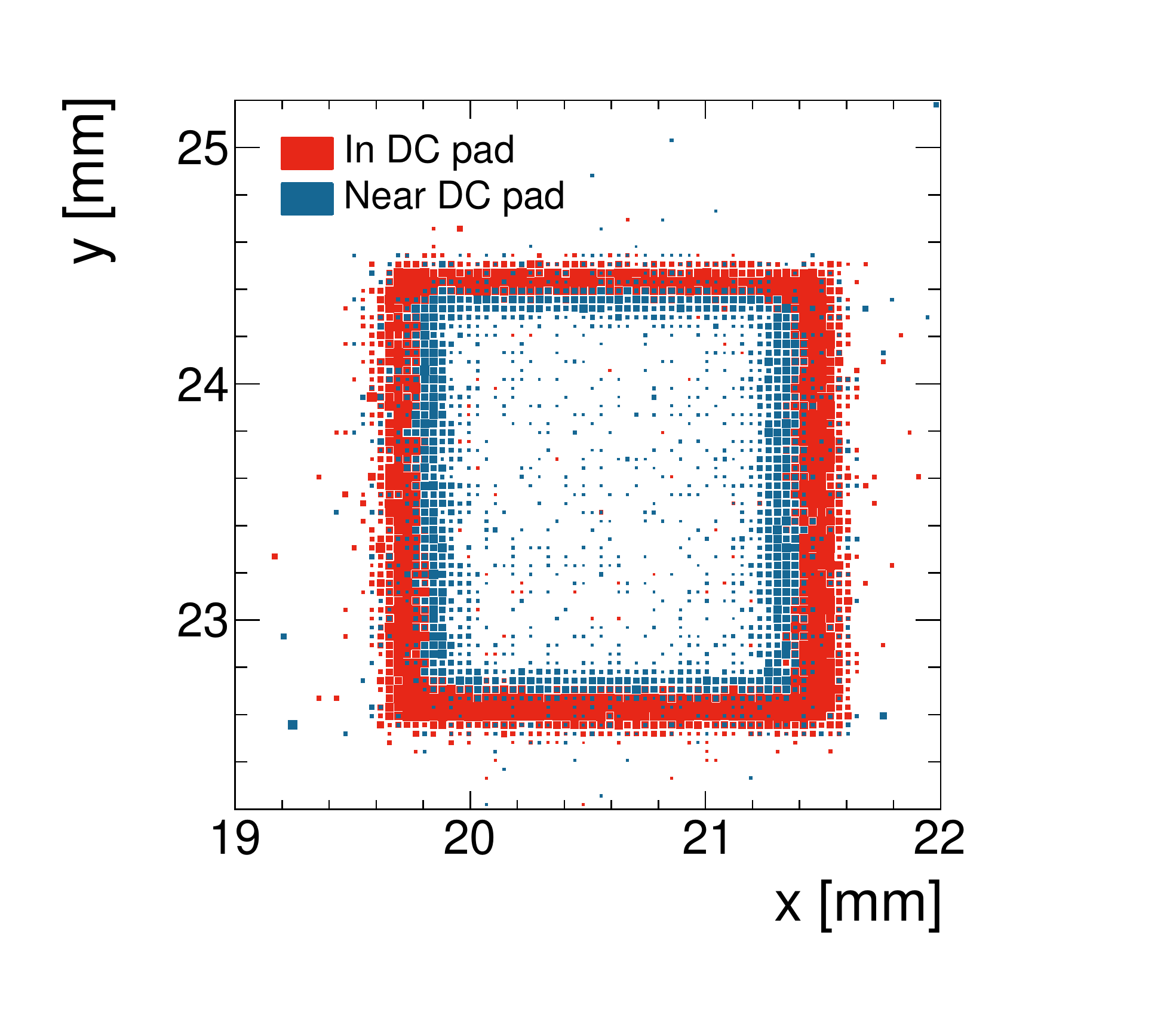}
        \caption{DC-pad signal populations versus proton position}\label{fig:dc_amplitude_position}
    \end{subfigure}
    \caption{(a) Charge collected by the DC pad. Data are shown in black, while a fit to data using a Landau function convoluted with a Gaussian function is shown in red. The most probable value (MPV) of the Landau is shown, along with the statistical uncertainty of the fit. (b) Hit position for different populations of DC-pad signals. Signals with amplitudes greater than \SI{30}{\milli \V} are shown in red, and have positions consistent with the DC pad itself. Lower amplitude signals (between $11$ and \SI{30}{\milli \V}) are shown in blue, and have positions in the active area of the sensor between strips and the DC pad.} 
    
\label{fig:dc_signals}
\end{figure}

As discussed in Section~\ref{sec:simulation}, when the DC contact is struck directly by a proton, the entire charge generated in the substrate is collected, and the generated DC-pad signal has the same characteristics as those of a standard unsegmented LGAD. We make use of this feature to estimate the gain of the sensor. As shown in Figure~\ref{fig:dc_landau}, we measure the charge collected by the DC pad to be $11~\rm{fC}$, with a $30\%$ systematic uncertainty in the calibration of the amplifier response to LGAD signals. This collected charge corresponds to a gain of approximately $17$.

As demonstrated in Figure~\ref{fig:dc_amplitude_position}, the amplitude of induced DC-pad signals decreases as the distance between the incident proton and the DC pad increases. This study was carried out by defining two different values of amplitude thresholds for the analysis of the DC-pad signals. Events with DC-pad signals which pass the higher threshold have proton positions consistent with a direct hit to the DC-pad active area. Events with lower amplitude DC-pad signals come from a population with proton positions which are slightly displaced from the DC pad, towards the inner part of the sensor. This lower amplitude threshold, \SI{11}{\milli \V}, is set as low as possible. The higher amplitude threshold is set to \SI{30}{\mV}, and is designed to select events in which a proton is directly incident to the DC pad.

\subsection{Efficiency measurement}

In order to measure the device efficiency, events are required to have a well measured proton track in the FTBF telescope, with a position consistent with the sensor active area. The detection efficiency is measured as the fraction of these events which also has a signal above threshold in an AC-LGAD strip.

For this measurement, the amplitude threshold is reduced from \SI{110}{\milli \V} to \SI{100}{\milli \V} in order to include signals produced by incident protons that generate low signal amplitudes. To ensure the smallest amplitude signals are from incident protons, and not from noise fluctuations, the timestamp of signal peak, $t_{\rm{peak}}$, relative to the Photek time reference is required to be consistent with that of an incident particle to within $\pm$\SI{2}{\nano \s}. 

\begin{figure}[h!]
    \centering
    \includegraphics[width=0.5\linewidth]{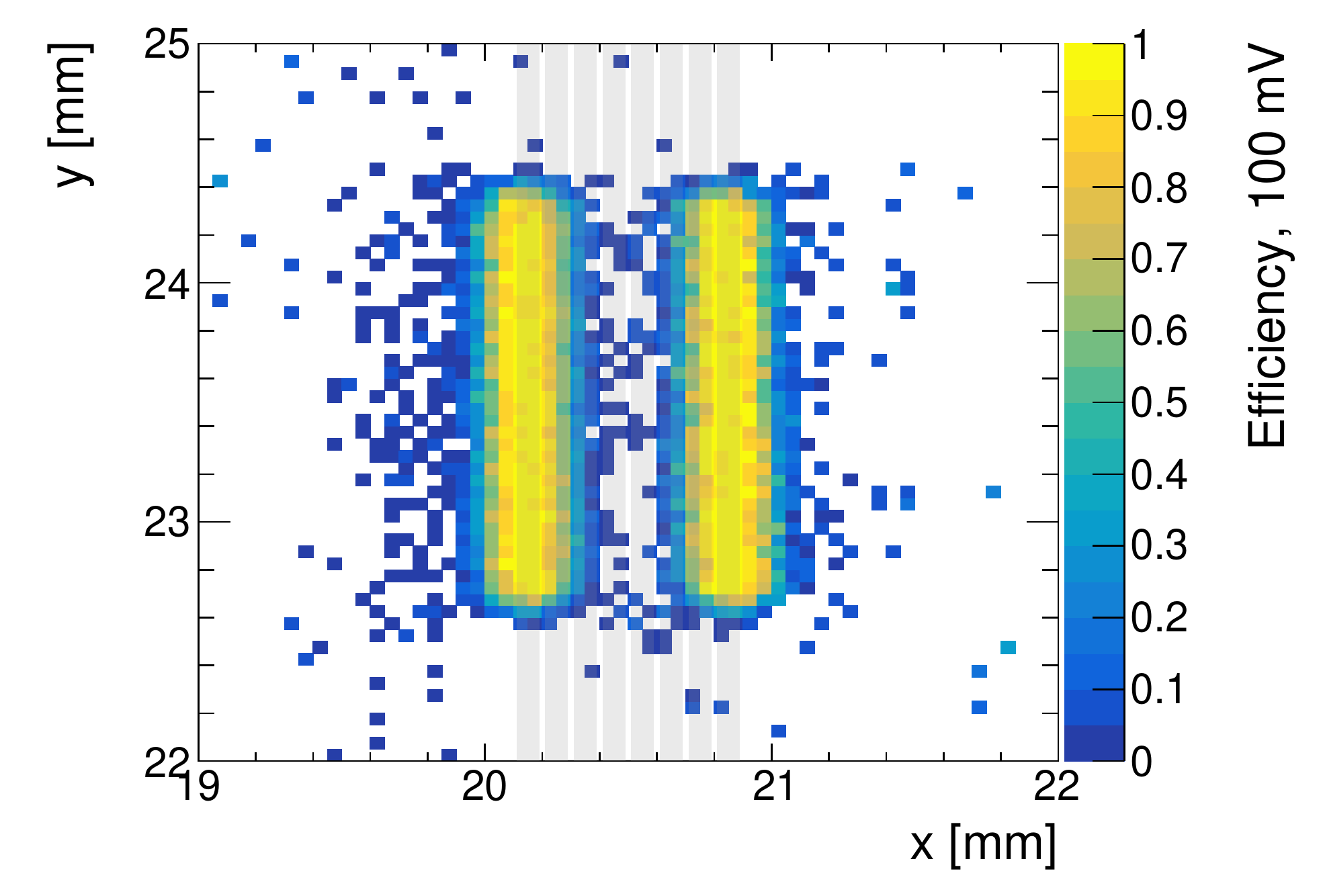}
    \caption{Signal efficiency for two read-out channels for signal amplitudes above a \SI{100}{\milli \V} threshold. The efficiency is shown as a function of the incident proton track positions along x and y directions. The hits on left-hand side correspond to the read-out channel 12, while those on right-hand side to channel 5, as defined in Fig.~\ref{fig:AC-LGAD_BOARD}. Shaded gray areas indicate the x-position of the metallized strip regions under study, as well as the strips in between. The efficiency from strips in between are not shown.  
    }
    \label{fig:sensor_efficiency_position}
\end{figure}

\begin{figure}[h!]
    \centering
     \includegraphics[width=0.8\linewidth]{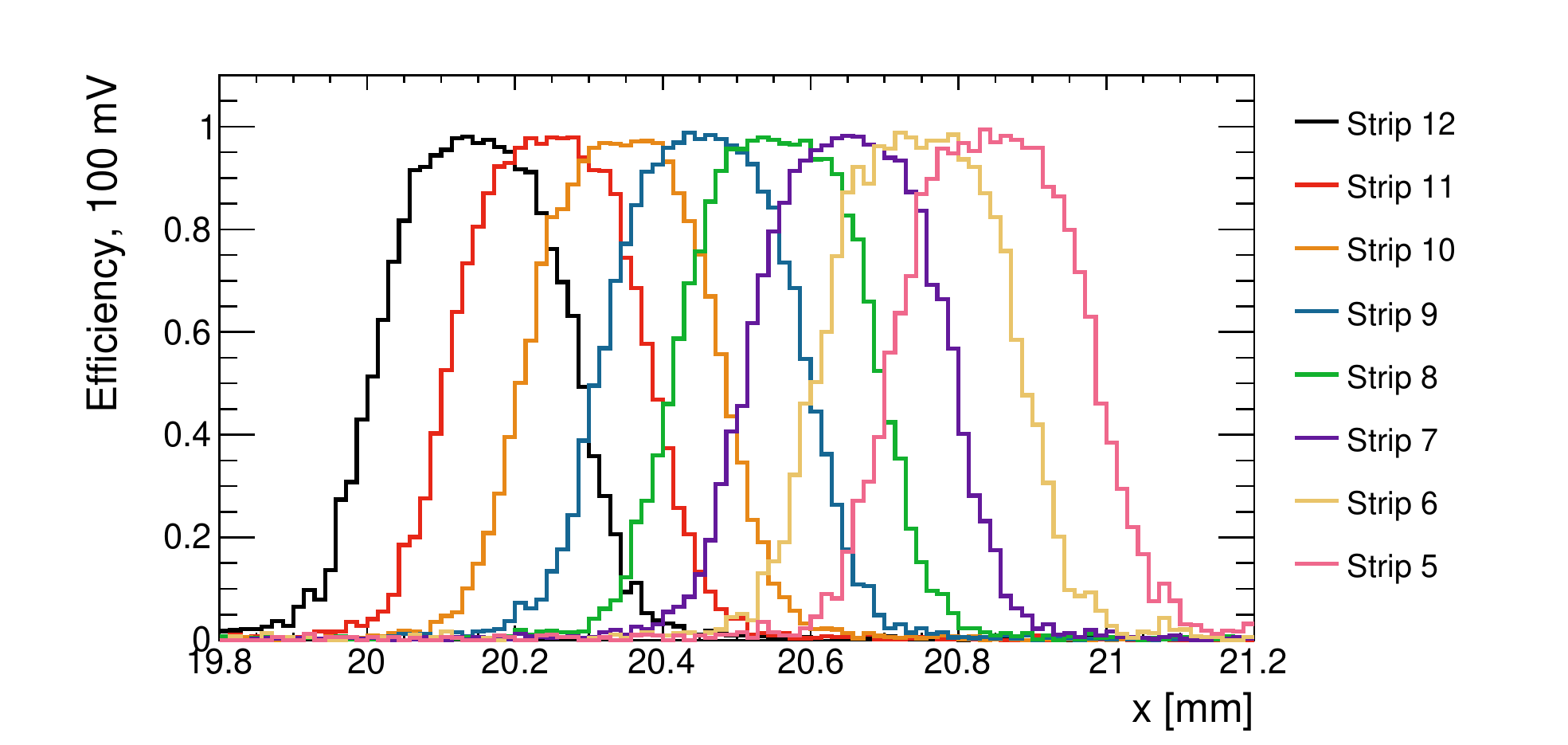}
    \caption{Efficiencies of individual strips as functions of incident proton x position. Strips are labeled by their position on the sensor, as illustrated in Fig.~\ref{fig:AC-LGAD_BOARD}. } \label{fig:strip_efficiency_v_x}
\end{figure}

Figure~\ref{fig:sensor_efficiency_position} shows the signal efficiency for two non-adjacent strips as a function of the incident proton's x and y position. Figure~\ref{fig:strip_efficiency_v_x} shows the efficiency for several individual strips as a function of the incident proton position in the x direction. Strips are labeled by their position on the sensor, as described in Fig.~\ref{fig:AC-LGAD_BOARD}. A selection on the track's y position, $22.9 < y <$ \SI{24.1}{\milli \m}, is made to ensure the proton is well-contained within the strips along the y-axis.  Similar efficiency profiles are observed for all strips, indicating good uniformity throughout the sensor area.

\begin{figure*}[t!]
    \centering
    \begin{subfigure}[t]{0.5\textwidth}
        \centering
        \includegraphics[width=0.95\textwidth]{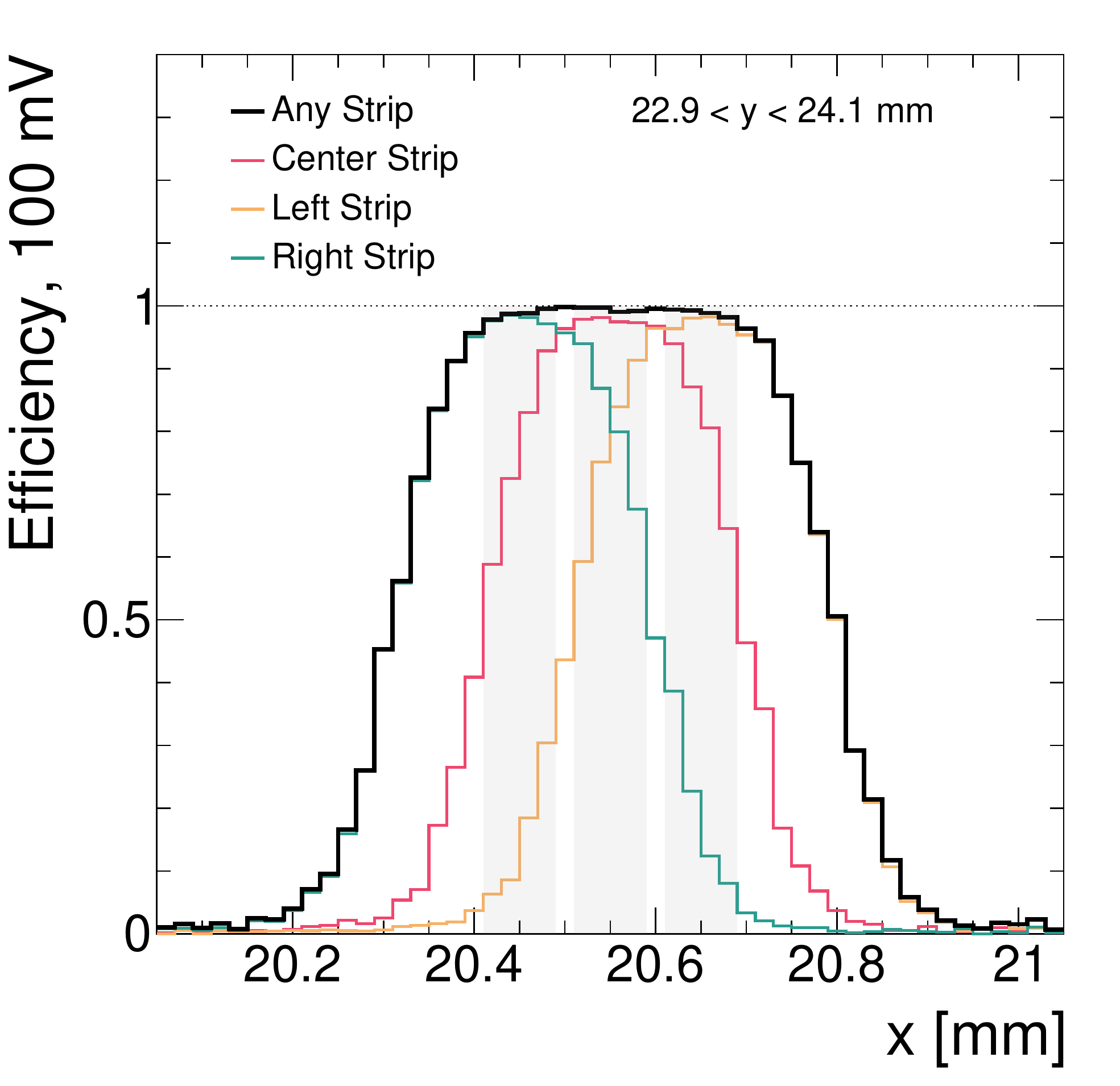}
        \caption{Combined efficiency in x direction}
    \end{subfigure}%
    ~ 
    \begin{subfigure}[t]{0.5\textwidth}
        \centering
        \includegraphics[width=0.95\textwidth]{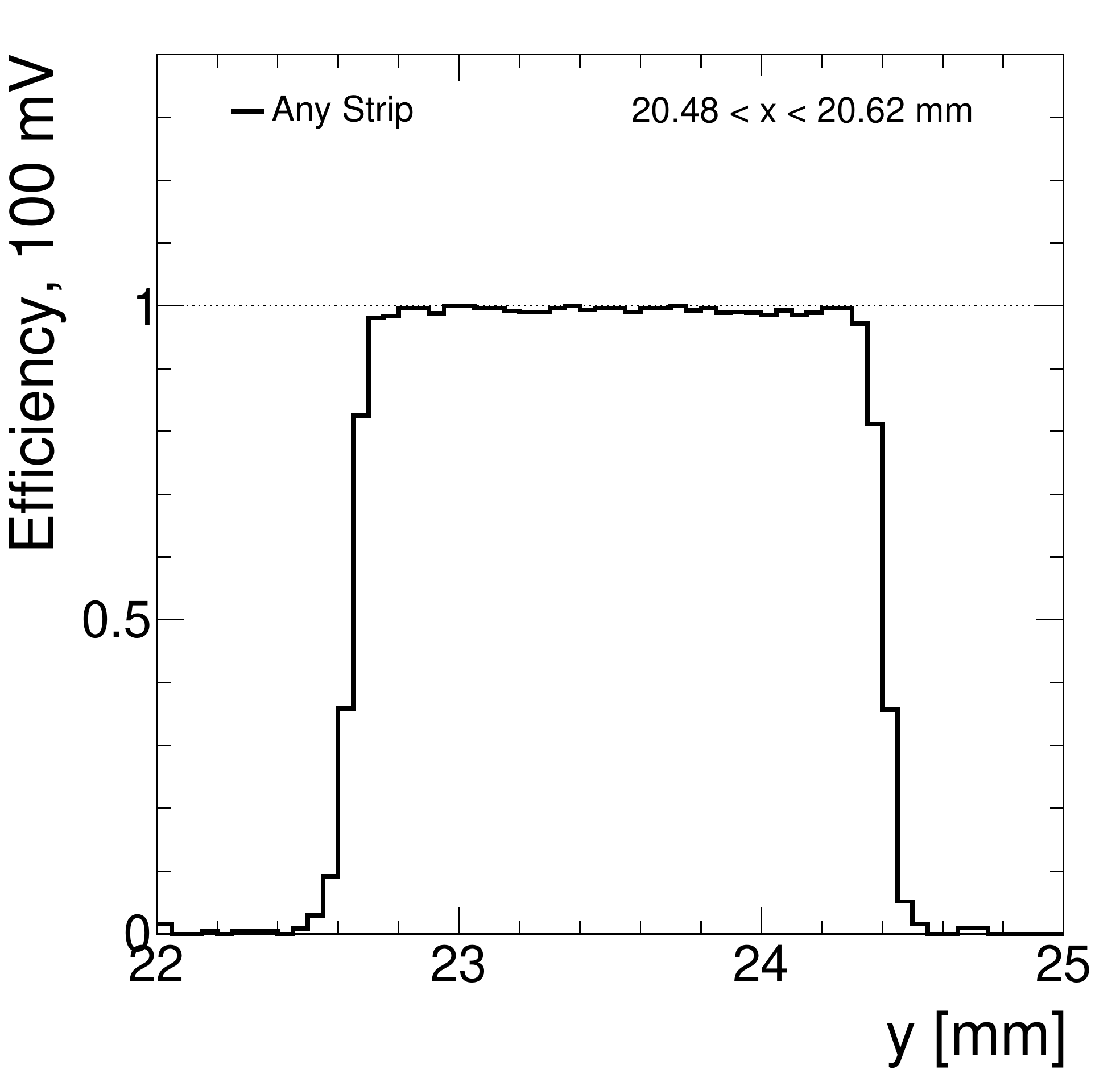}
        \caption{Combined efficiency in y direction}
    \end{subfigure}
    \caption{Combined signal efficiency of adjacent strips as a function of the proton track x and y position, for signals with amplitudes above a 100 mV threshold. In (a) individual strip efficiencies are also shown, and vertical grey bands indicate the strip positions in the x-direction.} \label{fig:total_efficiency}
\end{figure*}

Figure~\ref{fig:total_efficiency} shows the combined efficiency of adjacent strips in a cluster as a function of the incident proton position in x and y directions. When evaluating the sensor efficiency in the x direction, incident protons are required to have track positions in the range $22.9 < y < 24.1~\rm{mm}$. When evaluating the efficiency in the y direction, protons are required to have track positions in the range $20.48 < x < 20.62~\rm{mm}$. The combined efficiency is defined by requiring a hit in any of the three strips, with a threshold of \SI{100}{\milli \V}, and a $t_{\rm{peak}}$ consistent with an incident proton. The efficiencies of individual strips are also shown along the x direction for comparison. We measure a combined efficiency of $99.4\pm0.1\%$, and observe no evidence of efficiency loss between strips. 

The difference from an efficiency of $100\%$ can be attributed to protons with incorrectly measured positions, due to the FTBF telescope's finite spatial resolution of ${\sim}$\SI{50}{\micro \m}, and signals with amplitudes which are below the noise threshold.

\subsection{Measurements of spatial resolution}

The spatial resolution of AC-LGADs can be improved beyond the intrinsic strip resolution of $(\rm{strip~size})/\sqrt{12}$ by utilizing information about the relative signal amplitudes observed in strips within a cluster. Since most incident particles will produce a signal above threshold in at least two or three strips, we estimate that the position resolution of this sensor is on the order of \SI{10}{\micro\m} in the x-direction. 

By studying the residuals between the proton positions obtained from the FTBF telescope and from the AC-LGAD centroid reconstruction, we extract a spatial width of approximately \SI{50}{\micro\m}. This value is consistent with the spatial resolution of the FTBF telescope itself in this configuration, where the device under test is located far downstream from the telescope, with non-negligible material in between. As a result, we cannot make a precise measurement of the AC-LGAD spatial resolution, besides setting an upper bound of \SI{50}{\micro\m}. In future studies with an improved use of the tracker, we expect to gain sensitivity to resolutions closer to the intrinsic AC-LGAD potential.

\subsection{Measurements of time resolution}

\begin{figure*}[t!]
    \centering
    \includegraphics[width=0.5\textwidth]{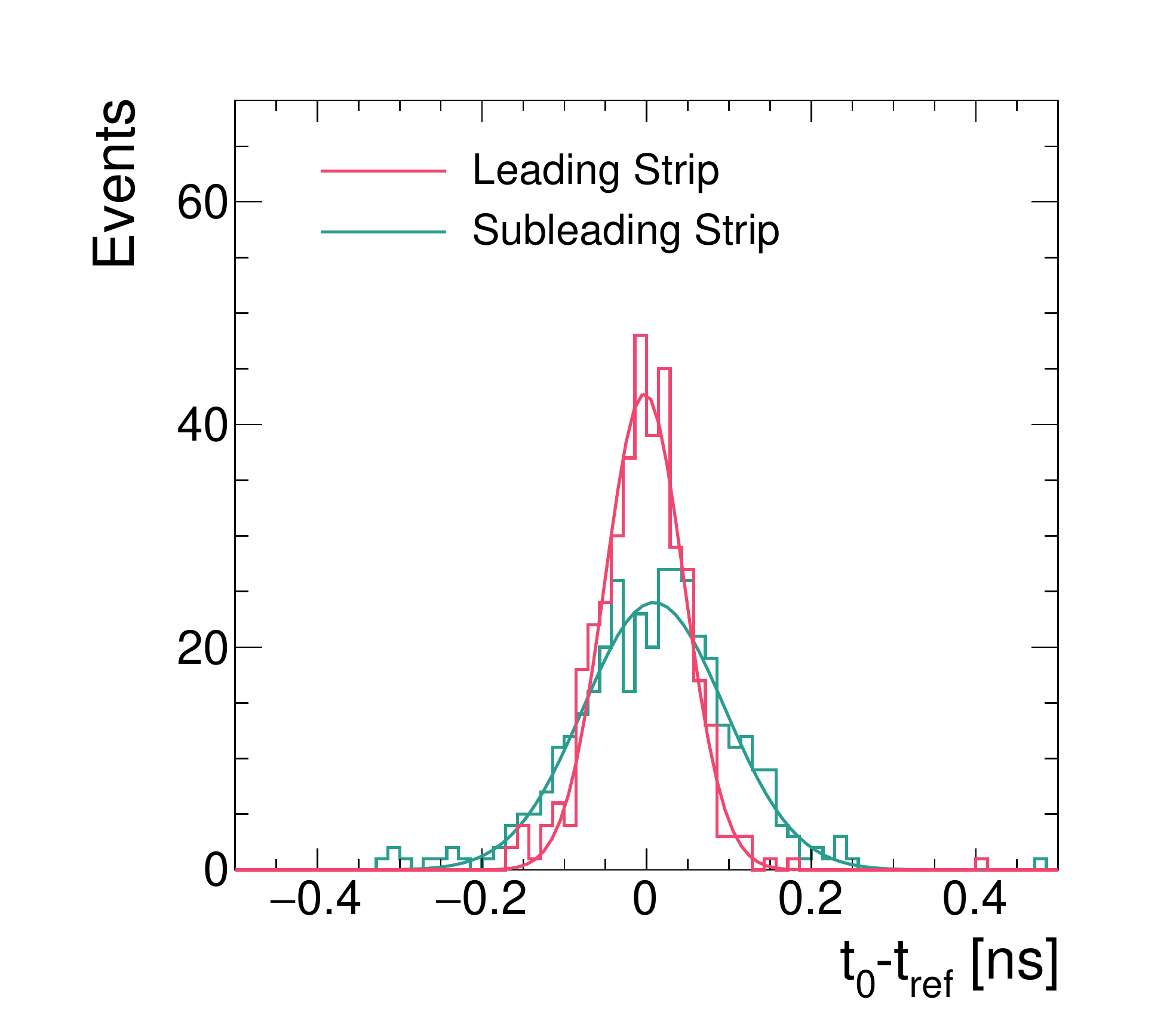}
    \caption{Time resolution measurement in events with a three-strip cluster. The calibrated time of arrival with respect to a reference time from the Photek device is shown for the leading, highest amplitude, and subleading, second highest amplitude, sensing channels. Each distribution is fit with a Gaussian function.} \label{fig:time_resolution}
\end{figure*}

The time resolution measurement for events with three strips above threshold is shown in Fig.~\ref{fig:time_resolution}. The time of arrival, $t_0$, is shown with respect to the Photek timestamp, $t_{\rm{ref}}$, which has a precision of \SI{10}{\pico\s}. The constant fraction timestamp is computed separately for each strip. Arrival times are calibrated for each channel, in order to account for differences in trace lengths on the readout board. The time of arrival is shown for the highest amplitude signal and for the next-to-highest amplitude signal.

Each distribution in Fig.~\ref{fig:time_resolution} is fit with a Gaussian function whose sigma is taken to be the time resolution. In events which have two or three strips with signals above threshold, we measure the time resolution to be on the order of \SI{45}-\SI{47}{\pico\s} for the strip with the highest amplitude. Neighboring strips have lower amplitude signals and correspondingly smaller slew rates, resulting in $70$-\SI{90}{\pico\s} time resolution. 

For this sensor, we do not observe significant improvement to the time resolution when combining measurements from multiple strips in three hit clusters. In future sensor designs we will study geometries with modified charge sharing among the neighboring strips in order to investigate if combining measurements can improve time resolution.

The intrinsic contribution to time resolution  arising from Landau ionization fluctuations for sensors of 50 $\mu$m active thickness is expected to be around \SI{30}{\pico \second}. This has been demonstrated with conventional, i.e. DC-coupled, LGADs read out with extremely low-noise amplifiers on single-channel readout board~\cite{Staiano_2017,apresyanStudiesUniformity502018,atlascollaborationLArHGTDPublicPlots,padilla2020effect}. The apparent difference between the measured resolution of about \SI{45}{\pico \second} and the \SI{30}{\pico \second} limit can be attributed to two factors. First, the 16-channel board used in this study produces additional noise that results in a non-negligible jitter contribution. Conventional LGADs read out by the 16-channel board have been measured to yield resolutions approximately $10$ -- \SI{15}{\pico \second} worse than those measured on a low-noise single-channel board~\cite{hellerStatusLGADCMS2019}. Second, due to the signal sharing effect in the capacitive coupling of the AC-LGAD, as reported in Sec. 5.1, the signals observed in the primary strip are reduced compared to conventional LGADs, which results in slightly poorer time resolution. Considering both effects, our observed resolution is consistent with the expectations for this configuration, and we see no evidence for additional contributions to the resolution. We expect that a similar sensor designed for doping concentrations yielding slightly larger gain values and read with lower noise electronics would reach the intrinsic \SI{30}{\pico \second} limit of conventional LGADs.

\section{Conclusions and Outlook}\label{sec:conclusions}

We exposed a $2 \times 2~\rm{mm}^2$ AC-LGAD strip sensor, fabricated at BNL, to the \SI{120}{\GeV} proton beam at the Fermilab Test Beam Facility (FTBF). The AC-LGAD sensor consists of an array of 17 AC-coupled metal strips with a pitch of \SI{100}{\micro\m} and a width of \SI{80}{\micro \m}, surrounded by a DC-connected pad and a Guard Ring. The reference position of incident protons was measured in a tracking system, while the reference timing by a Photek device. Pulses from adjacent strips were measured in the AC-LGAD sensor with a signal-to-noise ratio of about 27 for the highest amplitude pulses in an event. For the sum of hit amplitudes within in a cluster, the signal-to-noise ratio is ${\sim}47.5$. 
As TCAD numerical simulations predict, we observed that an incident proton induces current pulses on a few adjacent strips. In this particular sensor, most clusters are limited to three strips, and the signals are below threshold in strips farther away from the hit strip. The sum of signal amplitudes from a cluster is distributed according to a Landau function. The spatial resolution measured in this study is limited to \SI{50}{\micro \m}, due to the telescope spatial resolution at the beamline position of the AC-LGAD. However, it is expected that an interpolation of signal amplitudes will provide a spatial resolution significantly below \SI{50}{\micro \m} in the direction perpendicular to the strips. 
A hit efficiency close to $100\%$ has been measured for individual strips as well as for a combination of adjacent strips. It is therefore proven that this type of sensors allows for $100\%$ fill factor, as compared with standard LGAD technologies that show significant dead areas at the edges of the pixels. Time resolution on the order of $45$ -- \SI{47}{\pico\s} has been measured, compatible with LGADs operated with similar effective gain and read out by the same chain of electronics.
Future measurements will study the ultimate performance of such AC-LGAD sensors both in terms of spatial and time resolution, with more precise reference tracker resolution and higher sensor gain values as well as different geometrical sensor layouts.

\acknowledgments
We thank the FTBF personnel and Fermilab accelerator’s team for very good beam
conditions during our test beam time. We also appreciate the technical support of the
Fermilab SiDet department for the rapid production of wire-bonded and packaged LGAD
assemblies. We would like to thank T.~Nebel and J.~Wish for their technical support in the
experimental setup; L.~Uplegger for developing the telescope tracker and the DAQ system; and M.~Kiburg, E.~Niner and E.~Schmidt for ensuring smooth operations during our experiments. This document was prepared using the resources of the Fermi National Accelerator Laboratory (Fermilab), a U.S. Department of Energy, Office of Science, HEP User Facility. Fermilab is managed by Fermi Research Alliance, LLC (FRA), acting under Contract
No. DE-AC02-07CH11359. 

The work of W. Chen, G. D'Amen, G. Giacomini and A. Tricoli is supported by the U.S. Department of Energy under grant DE-SC0012704. This research used resources of the Center for Functional Nanomaterials, which is a U.S. DOE Office of Science Facility, at Brookhaven National Laboratory under Contract No. DE-SC0012704. The work of A.~Apresyan, K.~Di~Petrillo, and R.~Heller was supported by the KA25 Advanced Technology R\&D program of the U.S. Department of Energy. The work of H. Lee and C.-S. Moon has been supported by the National Research Foundation of Korea (NRF) grant funded by the Korea government (MSIT) (Grants No. 2018R1A6A1A06024970, No. 2020R1A2C1012322 and Contract NRF-2008-00460).

\bibliographystyle{report}
\bibliography{biblio}{}

\providecommand{\href}[2]{#2}\begingroup\raggedright\begin{thebibliography}{10}

\bibitem{GIACOMINI201952}
G.~Giacomini, W.~Chen, F.~Lanni, and A.~Tricoli, {\em {Development of a
  technology for the fabrication of Low-Gain Avalanche Diodes at BNL}\/},
  \href{http://dx.doi.org/https://doi.org/10.1016/j.nima.2019.04.073}{Nucl.
  Instrum. And Meth. in Phys. Res. A {\bf 934} (2019)  {52--57}}.
  \url{http://www.sciencedirect.com/science/article/pii/S0168900219305479}.

\bibitem{micronlgad}
N.~Moffat, R.~B. ans M.~Bullough, L.~Flores, D.~Maneuski, L.~Simon, N.~Tartoni,
  F.~Dohertya, and J.~Ashbya, {\em Low Gain Avalanche Detectors ({LGAD}) for
  particle physics and synchrotron applications\/},  JINST {\bf 13} (2018)
  C~03014.

\bibitem{hartmuth}
H.~F.-W. Sadrozinski, S.~Ely, V.~Fadeyev, Z.~Galloway, J.~Ngo, C.~Parker,
  B.~Petersen, A.~Seiden, A.~Zatserklyaniyv, N.~Cartiglia, F.~Marchetto,
  M.~Bruzzi, R.~Mori, M.~Scaringella, and A.~Vinattieri, {\em Ultra-fast
  silicon detectors\/},  Nucl. Instrum. And Meth. in Phys. Res. A {\bf 730}
  (2013)  226--231.

\bibitem{CMS:2667167}
{CMS Collaboration}, {\em {A MIP Timing Detector for the CMS Phase-2
  Upgrade}\/},   CERN-LHCC-2019-003. CMS-TDR-020, CERN, Geneva, Mar, 2019.
\newblock \url{https://cds.cern.ch/record/2667167}.

\bibitem{Collaboration:2623663}
{ATLAS Collaboration}, {\em {Technical Proposal: A High-Granularity Timing
  Detector for the ATLAS Phase-II Upgrade}\/},   CERN-LHCC-2018-023.
  LHCC-P-012, CERN, Geneva, Jun, 2018.
\newblock \url{http://cds.cern.ch/record/2623663}.

\bibitem{CERN-ACC-2015-0140}
I.~Bejar~Alonso and L.~Rossi, {\em {HiLumi LHC Technical Design Report:
  Deliverable: D1.10}\/},  \url{https://cds.cern.ch/record/206913}, Nov, 2015.

\bibitem{CERN-ATS-2012-236}
{The project is partially supported by the EC as FP7 HiLumi LHC Design Study
  under grant no. 284404 Collaboration}, L.~Rossi and O.~Bruning, {\em {High
  Luminosity Large Hadron Collider: A description for the European Strategy
  Preparatory Group}\/},  \url{https://cds.cern.ch/record/1471000}, Geneva,
  Aug, 2012.

\bibitem{RSD_NIM}
M.~Mandurrino, R.~Arcidiacono, M.~Boscardin, N.~Cartiglia, G.~F.~D. Betta,
  M.~Ferrero, F.~Ficorella, L.~Pancheri, G.~Paternoster, F.~Siviero, V.~Sola,
  A.~Staiano, and A.~Vignati, {\em Analysis and numerical design of Resistive
  AC-Coupled Silicon Detectors (RSD) for 4D particle tracking\/},
  \href{http://dx.doi.org/10.1016/j.nima.2020.163479}{Nucl. Inst. Meth. A {\bf
  959} (2020)  163479}.

\bibitem{WADA2019380}
S.~Wada et al., {\em {Evaluation of characteristics of Hamamatsu low-gain
  avalanche detectors}\/},
  \href{http://dx.doi.org/https://doi.org/10.1016/j.nima.2018.09.143}{Nucl.
  Instrum. And Meth. in Phys. Res. A {\bf 924} (2019)  380 -- 386}.
  \url{http://www.sciencedirect.com/science/article/pii/S0168900218312993}.
  11th International Hiroshima Symposium on Development and Application of
  Semiconductor Tracking Detectors.

\bibitem{DALLABETTA2015154}
G.-F.~D. Betta et al., {\em Design and {TCAD} simulation of double-sided
  pixelated low gain avalanche detectors\/},
  \href{http://dx.doi.org/https://doi.org/10.1016/j.nima.2015.03.039}{Nucl.
  Instrum. And Meth. in Phys. Res. A {\bf 796} (2015)  154 -- 157}. Proceedings
  of the 10th International Conference on Radiation Effects on Semiconductor
  Materials Detectors and Devices.

\bibitem{PELLEGRINI201624}
G.~Pellegrini et al., {\em {Recent technological developments on LGAD and iLGAD
  detectors for tracking and timing applications}\/},
  \href{http://dx.doi.org/https://doi.org/10.1016/j.nima.2016.05.066}{Nucl.
  Instrum. And Meth. in Phys. Res. A {\bf 831} (2016)  24 -- 28}.
  \url{http://www.sciencedirect.com/science/article/pii/S0168900216304557}.
  Proceedings of the 10th International Hiroshima Symposium on the Development
  and Application of Semiconductor Tracking Detectors.

\bibitem{ACLGADprocess}
G.~Giacomini, W.~Chen, G.~D'Amen, and A.~Tricoli, {\em Fabrication and
  Performance of AC-coupled LGADs\/},  JINST {\bf 14} (2019)  P09004.
  \url{https://doi.org/10.1088%2F1748-0221%2F14%2F09%2Fp09004}.

\bibitem{8846722}
M.~{Mandurrino}, R.~{Arcidiacono}, M.~{Boscardin}, N.~{Cartiglia}, G.~F. {Dalla
  Betta}, M.~{Ferrero}, F.~{Ficorella}, L.~{Pancheri}, G.~{Paternoster},
  F.~{Siviero}, and M.~{Tornago}, {\em Demonstration of 200-, 100-, and 50-
  $\mu$ m Pitch Resistive AC-Coupled Silicon Detectors (RSD) With 100$\%$
  Fill-Factor for 4D Particle Tracking\/},  IEEE Electron Device Letters {\bf
  40} (2019) no.~11, .

\bibitem{silvaco}
 \url{https://www.silvaco.com/products/tcad.html}.

\bibitem{FTBF}
{\em {Fermilab Test Beam Facility}\/},  \url{https://ftbf.fnal.gov}.

\bibitem{4775101}
M.~{Turqueti}, R.~A. {Rivera}, A.~{Prosser}, J.~{Andresen}, and
  J.~{Chramowicz}, {\em CAPTAN: A hardware architecture for integrated data
  acquisition, control, and analysis for detector development\/},  in {\em 2008
  IEEE Nuclear Science Symposium Conference Record}, pp.~3546--3552.
\newblock 2008.

\bibitem{RONZHIN2015288}
A.~Ronzhin, S.~Los, E.~Ramberg, A.~Apresyan, S.~Xie, M.~Spiropulu, and H.~Kim,
  {\em Study of the timing performance of micro-channel plate photomultiplier
  for use as an active layer in a shower maximum detector\/},
  \href{http://dx.doi.org/https://doi.org/10.1016/j.nima.2015.06.006}{Nucl.
  Instrum. Meth. A {\bf 795} (2015)  288 -- 292}.
  \url{http://www.sciencedirect.com/science/article/pii/S0168900215007500}.

\bibitem{Staiano_2017}
A.~Staiano, R.~Arcidiacono, M.~Boscardin, G.~D. Betta, N.~Cartiglia, F.~Cenna,
  M.~Ferrero, F.~Ficorella, M.~Mandurrino, M.~Obertino, L.~Pancheri,
  G.~Paternoster, and V.~Sola, {\em Development of Ultra-Fast Silicon Detectors
  for 4D tracking\/},
  \href{http://dx.doi.org/10.1088/1748-0221/12/12/c12012}{Journal of
  Instrumentation {\bf 12} (2017) no.~12, C12012--C12012}.
  \url{https://doi.org/10.1088%2F1748-0221%2F12%2F12%2Fc12012}.

\bibitem{apresyanStudiesUniformity502018}
A.~Apresyan, S.~Xie, C.~Pena, R.~Arcidiacono, N.~Cartiglia, M.~Carulla,
  G.~Derylo, M.~Ferrero, D.~Flores, P.~Freeman, Z.~Galloway, A.~Ghassemi,
  H.~Al~Ghoul, L.~Gray, S.~Hidalgo, S.~Kamada, S.~Los, M.~Mandurrino,
  A.~Merlos, N.~Minafra, G.~Pellegrini, D.~Quirion, A.~Ronzhin, C.~Royon,
  H.~Sadrozinski, A.~Seiden, V.~Sola, M.~Spiropulu, A.~Staiano, L.~Uplegger,
  K.~Yamamoto, and K.~Yamamura, {\em Studies of Uniformity of 50
  \si{\micro\meter} Low-Gain Avalanche Detectors at the {{Fermilab}} Test
  Beam\/},  \href{http://dx.doi.org/10.1016/j.nima.2018.03.074}{Nuclear
  Instruments and Methods in Physics Research Section A: Accelerators,
  Spectrometers, Detectors and Associated Equipment {\bf 895} (2018)
  158--172}.

\bibitem{atlascollaborationLArHGTDPublicPlots}
{ATLAS Collaboration}, {\em {{LArHGTD Public Plots}}\/},
  \url{https://twiki.cern.ch/twiki/pub/AtlasPublic/LArHGTDPublicPlots/tdr_timing_50_30um_CFD50-1.png}.
\newblock Note HPK-3.1-50 time resolution as function of gain.

\bibitem{padilla2020effect}
R.~Padilla, C.~Labitan, Z.~Galloway, C.~Gee, S.~M. Mazza, F.~McKinney-Martinez,
  H.~F.~W. Sadrozinski, A.~Seiden, B.~Schumm, M.~Wilder, Y.~Zhao, H.~Ren,
  Y.~Jin, M.~Lockerby, V.~Cindro, G.~Kramberger, I.~Mandiz, M.~Mikuz,
  M.~Zavrtanik, R.~Arcidiacono, N.~Cartiglia, M.~Ferrero, M.~Mandurrino,
  V.~Sola, and A.~Staiano, {\em Effect of deep gain layer and Carbon infusion
  on LGAD radiation hardness\/},  \url{https://arxiv.org/abs/2004.05260}, 2020.
\newblock \href{http://arxiv.org/abs/2004.05260}{{\tt arXiv:2004.05260
  [physics.ins-det]}}.

\bibitem{hellerStatusLGADCMS2019}
R.~Heller, {\em Status of {{LGAD R}}\&{{D}} for the {{CMS MIP Timing
  Detector}}\/},
  \url{https://indico.cern.ch/event/812761/contributions/3459059/}, June, 2019.
\newblock 34th RD50 Workshop.

\end{thebibliography}\endgroup
\end{document}